\documentclass[10pt,a4paper,twoside,twocolumn,american,prd,nofootinbib,superscriptaddress]{revtex4}
\usepackage{lmodern}

\usepackage[T1]{fontenc}
\usepackage[utf8]{inputenc}
\usepackage{graphicx}
\usepackage{color}
\usepackage{babel}
\usepackage{booktabs}
\usepackage{units}
\usepackage{amsmath}
\usepackage{amssymb}
\usepackage{esint}
\usepackage[unicode=true,pdfusetitle,
 bookmarks=true,bookmarksnumbered=false,bookmarksopen=false,
 breaklinks=false,pdfborder={0 0 0},backref=false,colorlinks=true]
 {hyperref}
\hypersetup{
 citecolor=blue,filecolor=blue,linkcolor=blue,urlcolor=blue}

\makeatletter  

 \@ifundefined{textcolor}{}
 {%
   \definecolor{BLACK}{gray}{0}
   \definecolor{WHITE}{gray}{1}
   \definecolor{RED}{rgb}{1,0,0}
   \definecolor{GREEN}{rgb}{0,1,0}
   \definecolor{BLUE}{rgb}{0,0,1}
   \definecolor{CYAN}{cmyk}{1,0,0,0}
   \definecolor{MAGENTA}{cmyk}{0,1,0,0}
   \definecolor{YELLOW}{cmyk}{0,0,1,0}
 }

\def\be{\begin{equation}}
\def\ee{\end{equation}}
\def\bea{\begin{eqnarray}}
\def\eea{\end{eqnarray}}

\newcommand{\vx}{{\mathbf x}}
\newcommand{\ct}{{\tau}} 

\newcommand{\HH}{{\mathcal H}}
\newcommand{\de}{\delta}
\newcommand{\ga}{\gamma}

\newcommand{\La}{\Lambda}

\makeatother 

\begin{document}

\title{General Relativistic $N$-body simulations in the weak field limit}

\author{Julian Adamek}
\email{julian.adamek@unige.ch}
\affiliation{D\'epartement de Physique Th\'eorique \& Center for Astroparticle Physics,
Universit\'e de Gen\`eve, Quai E.\ Ansermet 24, CH-1211 Gen\`eve 4, Switzerland}

\author{David Daverio}
\email{david.daverio@unige.ch}
\affiliation{D\'epartement de Physique Th\'eorique \& Center for Astroparticle Physics,
Universit\'e de Gen\`eve, Quai E.\ Ansermet 24, CH-1211 Gen\`eve 4, Switzerland}

\author{Ruth Durrer}
\email{ruth.durrer@unige.ch}
\affiliation{D\'epartement de Physique Th\'eorique \& Center for Astroparticle Physics,
Universit\'e de Gen\`eve, Quai E.\ Ansermet 24, CH-1211 Gen\`eve 4, Switzerland}

\author{Martin Kunz}
\email{martin.kunz@unige.ch}
\affiliation{D\'epartement de Physique Th\'eorique \& Center for Astroparticle Physics,
Universit\'e de Gen\`eve, Quai E.\ Ansermet 24, CH-1211 Gen\`eve 4, Switzerland}
\affiliation{African Institute for Mathematical Sciences, 6 Melrose Road, Muizenberg, 7945, South Africa}

\begin{abstract}
We develop a
formalism for General Relativistic $N$-body simulations in the weak field
regime, suitable for cosmological applications. The problem is kept
tractable by retaining the metric perturbations to first order, the first derivatives to second order and second derivatives
to all orders, thus taking into account the most important nonlinear effects of Einstein gravity. It is also expected that
any significant ``backreaction'' should appear at this order.
We show that the simulation scheme is feasible in practice by implementing it for a plane-symmetric situation and running two test
cases, one with only cold dark matter, and one which also includes a cosmological constant. For these plane-symmetric
situations, the deviations from the usual Newtonian $N$-body simulations remain small and, apart from a non-trivial correction
to the background, can be accurately estimated within the Newtonian framework. The correction to the background scale factor, which
is a genuine ``backreaction'' effect, can be robustly obtained with our algorithm. Our numerical
approach is also naturally
suited for the inclusion of extra relativistic fields and thus for dark energy or modified gravity simulations.
\end{abstract}
\maketitle

\section{Introduction}

Cosmological late-time calculations and simulations are usually divided into two distinct approaches, linear perturbation
theory of General Relativity (GR) and Newtonian $N$-body simulations. Linear perturbation theory is used on large scales 
where perturbations are small, so that higher-order terms can safely be neglected. On small scales on the other hand, 
Newtonian gravity provides a very good approximation to the dynamics of non-relativistic massive particles.

This distinction however breaks down in at least two, and maybe three cases: 

The first case concerns the addition of
relativistic fields in a context where nonlinearities are important. An example is the impact of topological defects on
the formation of structure \cite{Obradovic:2011mt} or the effect of modified-gravity (MG) theories 
on gravitational clustering on small scales, see~\cite{Brax:2012ie} and refs.\ therein. In this
case we have to simulate the nonlinear dynamics at least for the matter perturbations, and in general also for the fields,
as MG theories generically need nonlinear screening mechanisms that become important on small scales. Newtonian
gravity is usually not appropriate as e.g.\ both topological defects \cite{Magueijo:1996px,Durrer:1997ep} and MG theories \cite{Saltas:2010tt} exhibit a non-trivial anisotropic stress that
leads to a gravitational slip which cannot be included in standard $N$-body simulations. 

The second case is due to the ongoing revolution in observational cosmology where surveys are now reaching unprecedented
sizes, mapping out a significant fraction of the observable Universe. On large scales and at large distances it becomes necessary
to take into account relativistic effects \cite{Bonvin:2005ps,Bonvin:2011bg,Challinor:2011bk,Bertacca:2012tp,Umeh:2012pn,BenDayan:2012wi}.
Although small, the impact of the perturbations, e.g., on distance measurements, is not negligible and can either be used as an additional
probe of cosmology (e.g.\ \cite{Bonvin:2006en,Abate:2008au}) or lead to an additional noise in the measurements 
\cite{Valkenburg:2013qwa,BenDayan:2013gc} that may already be relevant for e.g.\ the Planck satellite results \cite{Marra:2013rba,Fleury:2013uqa}.
A general relativistic extension of $N$-body simulations will automatically include these effects, and since the metric
is fully known, this will in addition
allow us to integrate the geodesic equation of photons through the simulation volume to obtain accurate predictions
for observations that include all relevant relativistic effects.

If we implement a numerical scheme that is able to follow the relativistic evolution of the Universe, we are then also able
to test the third (and more speculative) case for a general relativistic framework for cosmological simulations
in the nonlinear regime: GR is a nonlinear theory, and in principle nonlinear effects on small scales
can ``leak'' to larger scales and lead to unexpected non-perturbative behavior. This idea is often called ``backreaction''
(see e.g.\ \cite{Rasanen:2006kp,Buchert:2007ik,Rasanen:2011ki,Clarkson:2011zq,Buchert:2011sx} for some recent reviews). If it is realized
in nature, it would link the recent onset of accelerated expansion to the beginning of nonlinear structure formation
and so provide a natural solution to the coincidence problem of dark energy. 
This cannot easily be checked within Newtonian simulations as the
terms from backreaction e.g. in the Buchert formalism~\cite{Buchert:2007ik} become total
derivatives in the Newtonian approximation and therefore do not contribute in a simulation with periodic boundary conditions~\cite{Rasanen:2010wz}.

However, it is very difficult
to test backreaction analytically due to the nonlinear nature of the problem, and a numerical GR simulation
appears to be the most straightforward
way to rigorously test this possibility. This test of backreaction is in a sense a bonus on top of the
important applications of such a code for precision cosmology in general and especially in the dark sector.

While it would be desirable to simulate cosmology in a fully general relativistic way, this is technically very challenging,
as demonstrated by the efforts necessary to simulate, for example, black hole mergers over just a short period of
time, see~\cite{lrr-2011-6} and refs.\ therein. We therefore need a scheme that captures the features of General
Relativity that are relevant for cosmology, without the overhead of using full GR. This can be done by first assuming
that, as in linear perturbation theory, our Universe on very large scales is close to being isotropic and homogeneous, i.e., has
a metric close to  the Friedmann-Lema\^\i tre-Robertson-Walker (FLRW) class and can thus be taken to have a line element
(keeping for the moment only the scalar perturbations, and assuming a flat FLRW background universe) of the form
\be
ds^2 = a^2(\ct) \left[ -(1+2\Psi) d\ct^2 + (1-2\Phi) d\vx^2 \right] \, . \label{eq:line1}
\ee
The success of linear perturbation theory for the analysis of perturbations in the cosmic microwave background (CMB) indicates that the gravitational potentials as well
as perturbations in the matter distribution like $\delta \doteq \delta\rho / \bar{\rho}$ are small on large scales. However, the
existence of galaxies, suns, planets and cosmological observers requires that at least the matter perturbations become
large on small scales, with $\delta\gg 1$. On small scales we expect that the matter perturbations and the gravitational
potential are connected to a high accuracy by Poisson's equation,
\be
\Delta \Phi = 4 \pi G a^2 \bar{\rho}~\delta\, . \label{eq:poisson}
\ee
In Fourier space $\Delta\Phi$ becomes $-k^2\Phi$, and on small scales the wavenumber $k$ is large, so that $\de$ can
become large
even if the potential $\Phi$  remains small on {\em all} scales.  Indeed, the potential is expected
to remain small since it is small initially and it does not grow within linear perturbation theory. 

The approach that we adopt is then to keep the gravitational
potentials always only to first order, but to be more careful with spatial derivatives. We keep first derivatives of $\Phi$ and $\Psi$ (and therefore also velocities)
to second order, and second and higher spatial derivatives of the potentials (and therefore also $\delta$) to all orders.
The theoretical foundations of this approach have been laid out recently in \cite{Green:2010qy,Green:2011wc}. Here, we want to take it an
important step further and develop the technology for its numerical application.
See also \cite{Rampf:2013ewa,Rigopoulos:2013nda} for a study of relativistic corrections to particle motion.

Notice that this weak field limit is enough to at least {\em test} for strong backreaction effects: as long as the metric perturbations
remain small, we know that we are within the domain of validity of our approximations. If however they become large, then our
scheme breaks down, but we learn in this case that the standard approaches to cosmological predictions break down as well.
This enables us to diagnose the failure in our approach, and to recognize whether we need to go beyond the weak-field limit.

In this paper we mainly describe our formalism and
the numerical algorithms, and perform tests in a plane symmetric situation. The application to realistic cosmological models in three spatial dimensions
and a detailed comparison with Newtonian $N$-body simulations will be presented in a forthcoming paper. An application
to observations in plane symmetric universes, and a comparison with exact relativistic solutions
is discussed in an accompanying paper~\cite{Adamek:2014qja}.

The outline of this paper is as follows: in the next section we describe our approximation scheme and present the basic equations.
In Section~\ref{sec:numerics} we outline the numerical implementation. In Section~\ref{sec:results_variables} we apply it to a simple,
plane-symmetric case.
We discuss the results and conclude in Section~\ref{sec:conclusions}.
Some details of the numerical implementation and
the generation of initial data are deferred to two appendices.

\section{Approximation scheme and fundamental equations\label{sec:scheme}}

When we go beyond linear perturbations, the line element (\ref{eq:line1}) does not allow for a self-consistent description
of cosmology, as scalar, vector and tensor perturbations start to mix. We therefore use in this paper the
more general line element which admits also vector and tensor modes,
\bea
ds^2 &=& a^2(\ct) \big[ -(1+2\Psi) d\ct^2 - 2 B_i dx^i d\ct + \nonumber \\
&& (1-2\Phi) \delta_{ij} dx^i dx^j  + h_{ij} dx^i dx^j \big] \, . \label{eq:metric}
\eea
In order to remove the gauge freedom we restrict $B_i$ to be a pure vector mode and $h_{ij}$ to be a pure tensor perturbation. To
this end we choose the transverse/traceless gauge conditions $\delta^{ij} B_{i,j} = \delta^{ik} h_{ij,k} = \delta^{ij} h_{ij} = 0$, as
was done in \cite{Green:2011wc}. The above line element and gauge conditions can be imposed without loss
of generality, i.e.\ we are not restricting the class of solutions, as long as we are interested in solutions which are not too far away
from FLRW and hence have a cosmological interpretation. On the other hand, even if the perturbations become large, e.g.\
$|\Psi| \gtrsim 1$, it is not clear that deviations from the Friedmann background are significant. It might just be that the longitudinal
gauge becomes badly adapted. This has been observed in some models of the early Universe, see e.g.~\cite{Brustein:1994kn,Cartier:2003jz},
but is not expected to occur in a situation which is close to Newtonian gravity. 

Having abandoned the Newtonian concept of absolute space, index placement has
become meaningful and must be treated accordingly. For convenience, and to avoid ambiguity or confusion, we will always consider the
perturbations $B_i$ and $h_{ij}$ with covariant indices only, and explicitly write the Euclidean metric wherever it is needed for index
contraction. We use the notation $f_{,i} \doteq \partial f/\partial x^i$, $\Delta f \doteq \delta^{ij} \partial^2 f / \partial x^i \partial x^j$,
and a prime ${}'$ to denote $\partial/\partial \tau$. Latin indices take values $1, 2, 3$, while Greek indices run from $0$ to $3$,
and a sum is implied over repeated indices.

Except for very particular, symmetric cases, solving Einstein's equations can only be achieved in approximation. The approximation
scheme we use here is well adapted for cosmological settings where one is interested in the solution far away from compact sources
(like black holes) and one can therefore consider a weak field limit. However, we clearly want to go beyond linear perturbation
theory in order to allow for matter perturbations to evolve fully into the nonlinear regime of structure formation. We will therefore
use an approach that is equivalent to the formalism of \cite{Green:2010qy,Green:2011wc} who have adapted a shortwave approximation 
in order to treat small scale inhomogeneities in
cosmology. In this approach it is assumed that large (nonlinear) matter perturbations generally occur on small spatial
scales, an assumption that is well supported by observations. Accordingly, spatial derivatives of metric perturbations
should have a larger weight in a perturbative expansion than the metric perturbations themselves. In fact, gravitational potentials
are of the order of $\sim 10^{-5}$ from galactic scales out to the scales of CMB observations. Gradients, on the other hand, need not
be that small. They are related to peculiar velocities, and these are typically observed to be of the order of $\sim 10^{-3}$ on
galactic and cluster scales, see~\cite{Turnbull:2011ty} and references therein. Second spatial derivatives are related to the density
contrast, which becomes non-perturbative, i.e.\ larger than unity.

In more detail, the approximation scheme\footnote{We present the scheme from a slightly different point of view than
\cite{Green:2011wc}, whose approach may be conceptually more elaborate but also more difficult to grasp. In particular, the weight
of each term in the perturbative expansion is considered separately at small and large scales. The reader who is interested
in a more detailed derivation of the equations is invited to study their accounts.}
we want to implement amounts to giving every spatial derivative a weight $\epsilon^{-1/2}$
where $\epsilon$ is a weight characterizing the smallness of the metric perturbations. We then consistently keep all the terms up
to order $\epsilon$. This means that we keep terms
which are linear in the metric perturbations either by themselves or multiplied with second spatial derivatives which are considered
to be of $\mathcal{O}(1)$, so that we are accurate to $\mathcal{O}(\epsilon)$. In
addition, we retain quadratic terms of first spatial derivatives.
As velocities are expected to be of the same order as gradients, we go to quadratic order also here
to remain self-consistent.  Correspondingly, we have to keep density fluctuations, which are of  $\mathcal{O}(1)$
multiplied with metric perturbations or with squared velocities and squared spatial derivatives of metric perturbations.
Time derivatives are simply given the same weight
as the quantity they act upon.
This means, however, that we do not take a quasi-static limit where time derivatives are neglected altogether.

On very large scales, where all perturbations remain small, the scheme is equivalent to relativistic linear
perturbation theory since we do not, for instance, keep terms which are quadratic in the metric perturbations themselves.
On these scales, linear perturbation theory is very accurate because second-order corrections are expected to be
of the order of $\sim 10^{-10}$.
On intermediate and small scales, however, the scheme takes into account also the most important nonlinear relativistic terms.
Since we stay in a relativistic framework throughout, we ensure that gauge issues are automatically addressed in the correct way
such that we could easily construct all kinds of gauge invariant observables consistently. Such observables are, however, not the
focus of this paper, which is aimed at the development of the necessary tools.

With the metric~(\ref{eq:metric}) and our choice of gauge, the components of the Einstein tensor read
\begin{subequations}
\bea
G_0^0 &+& 3 \frac{\mathcal{H}^2}{a^2} = \frac{2}{a^2} \bigg[3 \mathcal{H} \Phi' + 3 \mathcal{H}^2 \Psi \nonumber \\ &&
- \left(1 + 4 \Phi\right) \Delta \Phi - \frac{3}{2} \delta^{ij} \Phi_{,i} \Phi_{,j}\bigg] \, , \label{eq:G00} \\
G_i^0  &=& -\frac{2}{a^2} \bigg[\frac{1}{4} \Delta B_i + \Phi'_{,i} + \mathcal{H} \Psi_{,i}\bigg] \, , \label{eq:G0i} \\
G_j^i &+& \delta_j^i \left(2\frac{\mathcal{H}'}{a^2}+\frac{\mathcal{H}^2}{a^2} \right) = \frac{\delta_j^i}{a^2} \bigg[2 \Phi''
+ 4 \mathcal{H} \Phi' + 2 \mathcal{H} \Psi' \nonumber \\ && + 2 \Psi \left(2 \mathcal{H}' + \mathcal{H}^2\right)
- \left(1 + 4 \Phi\right) \Delta \Phi - 2 \delta^{mn} \Phi_{,m} \Phi_{,n} \phantom{\bigg[} \nonumber \\
&&+ \left(1 + 2 \Phi - 2 \Psi\right) \Delta \Psi - \delta^{mn} \Psi_{,m} \Psi_{,n} \bigg] \nonumber \\
&&+ \frac{\delta^{ik}}{a^2} \bigg[\left(1 + 4 \Phi\right) \Phi_{,jk} - \left(1 + 2 \Phi - 2 \Psi\right)
\Psi_{,jk} \nonumber \\ && + B'_{\left(j,k\right)} + 2 \mathcal{H} B_{\left(j,k\right)} + h_{jk}'' + 2 \mathcal{H} h'_{jk}
- \Delta h_{jk} \phantom{\bigg]} \nonumber \\ && \hspace{35pt} + \Psi_{,j} \Psi_{,k} - 2 \Psi_{\left(,j\right.} \Phi_{\left.,k\right)}
+ 3 \Phi_{,j} \Phi_{,k}\bigg] \, , \label{eq:Gij}
\eea
\end{subequations}
up to terms which are higher order corrections to our approximation scheme (cf.\ Appendix~A of \cite{Green:2011wc}).
Here $\HH(\tau)=H(\tau)a(\tau)=a'/a$ is the comoving Hubble rate. For the time-time
component~(\ref{eq:G00}) and the spatial trace of~(\ref{eq:Gij}), the FLRW background terms
were moved to the left hand side. Indices in parentheses are to be symmetrized.

The metric perturbations are sourced by the perturbations of the total stress-energy tensor $T_\mu^\nu$. It contains contributions
from all types of matter, possibly a cosmological constant, and any other additional components one wants to include in a model.
We will explicitly consider the contributions from non-relativistic matter (baryons and cold dark matter) and a cosmological constant
$\Lambda$, and keep an unspecified remainder to collectively account for other sources of stress-energy:
\be
T_\mu^\nu = {T_\mathrm{m}}_\mu^\nu - \frac{\Lambda}{8 \pi G} \delta_\mu^\nu + {T_\mathrm{X}}_\mu^\nu\,.
\ee
We do not distinguish between baryons and cold dark matter.
In a perturbed universe, physical quantities like e.g.\ the energy density of matter are affected by the metric perturbations.
Our aim is to take these corrections consistently into account when running a numerical simulation of structure formation. For
convenience, we will therefore write the components of the matter stress-energy tensor in terms of ``bare'' quantities which can easily be 
computed in any standard $N$-body framework. In these frameworks, the phase space distribution of dark matter particles is usually
sampled by a collection of test particles whose positions and peculiar velocities are followed through a simulation. By making
a particle-to-mesh projection, it is possible to obtain average quantities like the (bare) bulk velocity
$u^i \doteq \langle v^i \rangle = \langle \partial x^i / \partial \tau \rangle$,
where $\langle \cdot \rangle$ denotes the averaging procedure associated with the projection method.
The method consists of a prescription of how to assign $N$-body particle properties, like mass or velocity, to nearby grid points,
thus constructing a matter density and velocity field on the grid by means of a weighted average, see Appendix~\ref{app:p2m}
for more details. This can be considered as ``the poor mans equivalent'' of a phase space integral.

We assume here that this procedure
is carried out in the standard way without any knowledge about the perturbed metric, which is why we refer to ``bare''
quantities. Therefore, the projection gives rise to the bare rest mass density defined as
\be
\rho \doteq \frac{\mathrm{rest~mass}}{\mathrm{coordinate~volume}} \times a^{-3}~.
\ee
In terms of the bare quantities, the physical (dressed) energy density of matter reads
\be
\rho_\mathrm{phys} \doteq -{T_\mathrm{m}}_0^0 = \left[1 + 3 \Phi + \frac{1}{2} \langle v^2 \rangle\right] \rho \, , \label{eq:T00}
\ee
up to higher order corrections which we neglect. As one can see, our approximation scheme takes into account the leading corrections
coming from the perturbation of the volume and the kinetic energy of the particles.

It should also be noted that even the homogeneous modes of $\rho$ and $\rho_\mathrm{phys}$ do not agree in general, because the
kinetic energy is strictly positive and hence has a positive average over the hypersurface of constant time. It is therefore necessary
to specify more precisely how one wants to perform the split between background and perturbations. We decide to evolve the scale factor
according to Friedmann's equation using an exactly pressureless equation of state for cold dark matter (CDM), in the spirit of Newtonian
$N$-body simulations. This means that we treat all finite velocity effects entirely on the level of the perturbations. As we shall see
later, this leads to a non-vanishing homogeneous mode in $\Phi$, which accounts for the correction to the scale 
factor\footnote{A homogeneous mode in $\Psi$ on the other hand can be gauged away by a global reparametrization of time. We will take the conventional definition of
cosmic time by imposing that $\Psi$ has a vanishing homogeneous mode.} due to an effective pressure
and other quadratic contributions. In fact, this correction can be interpreted as a ``backreaction'' term. However, as long as the
homogeneous mode remains small, the scale factor $a$ still gives a meaningful description of the background cosmology.

In short, for conceptual simplicity, we will use the ``bare'' $\rho$ and $\delta$ which one would infer from a given particle configuration
assuming an unperturbed Friedmann model with scale factor $a$, but we have to keep in mind that these quantities have to be interpreted
differently in the perturbed universe and have to be dressed with appropriate corrections when they enter the perturbed Einstein equations.

From $G_0^0 = 8 \pi G T_0^0$, and using Friedmann's equation, we obtain a first evolution equation for the metric:
\begin{multline}
\left(1 + 4 \Phi\right) \Delta \Phi - 3 \mathcal{H} \Phi' - 3 \mathcal{H}^2 \Psi +
\frac{3}{2} \delta^{ij} \Phi_{,i} \Phi_{,j} \\ = 4 \pi G a^2 \bar{\rho}
\left[\delta + 3 \Phi \left(1 + \delta\right) + \frac{1}{2} \left(1 + \delta\right) \langle v^2 \rangle\right]
\\- 4 \pi G a^2 \delta{T_\mathrm{X}}_0^0 \, . \label{eq:phi}
\end{multline}
Here, $\delta{T_\mathrm{X}}_0^0 \doteq {T_\mathrm{X}}_0^0 - {\overline{T}_\mathrm{X}}_0^0$, where
$-{\overline{T}_\mathrm{X}}_0^0$ and $\bar{\rho}$ are the homogeneous modes of the energy densities as they occur
in Friedmann's equation. In particular,
\be
4 \pi G \bar{\rho} = \frac{3}{2} H_0^2 \Omega_\mathrm{m} \left(1 + z\right)^3 \, ,
\ee
with $H_0$, $\Omega_\mathrm{m}$ and $z$ being bare parameters of the FLRW background. The Poisson equation (\ref{eq:poisson})
is obtained from Eq.~(\ref{eq:phi}) by dropping all terms except the first one on each side. 

One can view Eq.~(\ref{eq:phi})
as a parabolic (diffusion-type) equation for $\Phi$ to which Eq.~(\ref{eq:poisson}) is a first approximation if the diffusion timescale
$t_\mathrm{diff}$ is much shorter than the dynamical timescale $t_\mathrm{dyn}$ of the source term. To illustrate this statement,
let us consider a structure of size $r$ consisting of non-relativistic matter. A quick estimate gives\footnote{As diffusion is a random process,
the diffusion length is proportional to $\sqrt{\kappa t}$ where $\kappa = a/(3 \mathcal{H})$ is the diffusion coefficient that can be read off
from the prefactors of $\Delta\Phi$ and $\dot{\Phi}= a\Phi'$ in Eq.\ (\ref{eq:phi}). The diffusion time scale for a structure of size $r$ is then 
$t_{\rm diff} = r^2/\kappa$.}
\be
\label{eq:diffusiontime}
\frac{t_\mathrm{diff}}{t_\mathrm{dyn}} \simeq \frac{r^2}{r_H^2} \sqrt{1 + \delta} \, ,
\ee
where $r_H = a \mathcal{H}^{-1}$ is the Hubble radius, and we use the free-fall time $\simeq r_H/\sqrt{1+\delta}$ 
as an indicator for $t_\mathrm{dyn}$.

This ratio is tiny for any realistic structure with $r \ll r_H$. For instance, plugging in typical numbers for a stellar object
yields $t_\mathrm{diff} / t_\mathrm{dyn} \sim 10^{-20}$,
while for a structure comparable to the local supercluster one obtains
$t_\mathrm{diff} / t_\mathrm{dyn} \sim 10^{-5}$. This means that $\Phi$ will simply adjust adiabatically to its ``instantaneous''
equilibrium solution on these scales, which is why Eq.~(\ref{eq:poisson}) is a reasonable approximation in this case. On cosmological
scales $r \simeq r_H$, however, one may expect that retardation effects may become relevant. Furthermore, gauge dependence
typically becomes an issue as well on these scales, even in the linear regime. In our case this leads, for instance, to the
$\mathcal{H}^2 \Psi$-term present in Eq.~(\ref{eq:phi}) which does not appear in Poisson's equation.

In order to obtain a second scalar equation, it is useful to combine the time-time components and spatial trace in the following way:
\be
\frac{a^2}{2} \left(G_i^i - 3 G_0^0 - \frac{1}{\mathcal{H}} {G_0^0}'\right) = 4 \pi G a^2 \left(T^i_i - 3 T^0_0 -
\frac{1}{\mathcal{H}} {T_0^0}'\right)
\ee
Since $T_\mu^\nu$ is covariantly conserved, one can replace ${T_0^0}'$ essentially by spatial derivatives of the stress-energy
tensor, which are more easily computed at every time step of a simulation. After some algebra and using again Friedmann's
equations, one arrives at
\begin{multline}
\left(1 + 2 \Phi - 2 \Psi\right) \Delta \Psi - \delta^{ij} \Psi_{,i} \left(\Phi_{,j} + \Psi_{,j}\right) \\
+ \frac{1}{\mathcal{H}} \left(\Delta \Phi' + 4 \Phi \Delta \Phi' + 4 \Phi' \Delta \Phi\right) +
\frac{3}{\mathcal{H}} \delta^{ij} \Phi_{,i}' \Phi_{,j} \\= 4 \pi G \frac{a^2}{\mathcal{H}} \bigg[
-\bar{\rho} \left(1 + 3 \Phi\right) \left(\left(1 + \delta\right) \langle \gamma v^i \rangle\right)_{,i}
- \bar{\rho} \left(1 + \delta\right) u^i \Psi_{,i}\\ + 3 \bar{\rho} \Phi' \delta + \delta{T_\mathrm{X}}_{0,i}^i
- \delta{T_\mathrm{X}}_0^i \left(3 \Phi_{,i} - \Psi_{,i} + B'_i\right)\phantom{\bigg]}\\ - \Phi' \left(3 \delta{T_\mathrm{X}}_0^0 -
\delta{T_\mathrm{X}}_i^i\right) - \frac{1}{2} \delta^{ik} h'_{jk}  \delta{T_\mathrm{X}}_i^j\bigg] \, , \label{eq:psi}
\end{multline}
up to terms which we neglect in our approximation scheme. Here, because a spatial derivative acts on it, we have to use the
relativistic momentum up to cubic order in velocity,
i.e.\ we take into account the Lorentz factor which we approximate to quadratic order,
$\ga \simeq 1+v^2/2$. Dropping again all terms except the
first one on either side gives a Poisson equation for $\Psi$ which, however, is sourced by the divergence of the momentum density
instead of the density perturbation which we had in Eq.~(\ref{eq:poisson}). Of course these two are related by the continuity equation,
and hence the two expressions are the same to a first approximation, e.g.\ for linear perturbations of CDM.

In Eqs.~(\ref{eq:phi}) and (\ref{eq:psi}) we make no assumptions about the size of the perturbations
$\delta{T_\mathrm{X}}_\mu^\nu$, so they can be used to obtain the scalar metric
perturbations $\Phi, \Psi$ in any setting where these stress-energy terms can be reliably computed, whatever their origin may be.

Note also that Eq.~(\ref{eq:phi}) is a purely scalar equation. Moreover, whenever the two terms $\delta{T_\mathrm{X}}_0^i B'_i$ and
$\delta^{ik} h'_{jk}  \delta{T_\mathrm{X}}_i^j$ can be neglected, also Eq.~(\ref{eq:psi}) decouples from vector and tensor
perturbations.  In a numerical simulation, one can then solve these equations first and use the solutions in order to solve for the
vector and tensor quantities separately. This is the approach we will follow in Section~\ref{sec:numerics}, where we discuss the
implementation of this scheme for a pure $(\Lambda)$CDM universe (where $\delta{T_\mathrm{X}}_\mu^\nu = 0$). The equations are,
of course, still coupled indirectly by the evolution of $T_\mu^\nu$.

The equation for vector perturbations is
\begin{multline}
 -\frac{1}{4} \Delta B_i - \Phi'_{,i} - \mathcal{H} \Psi_{,i} \\= 4 \pi G a^2 \bigg[\bar{\rho} \left(1 + \delta\right)
\left(\delta_{ij} u^j - B_i\right) + \delta{T_\mathrm{X}}_i^0\bigg] \, . \label{eq:B}
\end{multline}
Note that the left hand side is already written in Helmholtz decomposition and that the longitudinal component of
the equation is completely determined by $\Phi$ and $\Psi$. On the other hand, by taking the curl of it, we could get rid of
the scalar potentials.
However, the nonlinear term $\de\cdot u^j$ remains and the term $\de\cdot B_i$ prevents us from writing an equation for
the curl of $B_i$ only. Therefore, we prefer to solve Eq.~(\ref{eq:B}) directly
after subtracting its longitudinal component through the use of $\Phi$ and $\Psi$ (see also Section~\ref{sec:vec_imp} for more
details on how we do this in practice).

Finally, in order to obtain an equation for the tensor perturbations, it is useful to consider the traceless combination
\be
G_j^i - \frac{1}{3} \delta_j^i G_k^k = 8 \pi G \left(T_j^i - \frac{1}{3} \delta_j^i T_k^k\right) \, ,
\ee
which yields\footnote{In equations
(A2) and (A4) of \cite{Green:2011wc} they include a contribution from $B_i$ on the right hand side, which to our understanding
should be dropped because it is multiplied by a velocity and hence is of higher order in the approximation scheme.}
\begin{multline}
 h''_{ij} + 2 \mathcal{H} h'_{ij} - \Delta h_{ij} + B'_{\left(i,j\right)} + 2 \mathcal{H} B_{\left(i,j\right)}
 + \left(1 + 4 \Phi\right) \Phi_{,ij} \\ - \left(1 + 2 \Phi - 2 \Psi\right) \Psi_{,ij} + \Psi_{,i} \Psi_{,j}
 - 2 \Phi_{\left(,i\right.} \Psi_{\left.,j\right)} + 3 \Phi_{,i} \Phi_{,j} \phantom{\bigg]} \\
 - \frac{1}{3} \delta_{ij} \bigg[\left(1 + 4 \Phi\right) \Delta \Phi - \left(1 + 2 \Phi - 2 \Psi\right) \Delta \Psi
 + \delta^{mn} \Psi_{,m} \Psi_{,n} \\ - 2 \delta^{mn} \Psi_{,m} \Phi_{,n} + 3 \delta^{mn} \Phi_{,m} \Phi_{,n} \bigg] =
 8 \pi G a^2 \bigg[\bar{\rho} \left(1 + \delta\right) \pi_{ij}\\ + \delta_{ik} \delta{T_\mathrm{X}}_j^k -
\frac{1}{3} \delta_{ij}  \delta{T_\mathrm{X}}_k^k\bigg] \, . \label{eq:h}
\end{multline}
Here, $\pi_{ij}$ is the bare anisotropic stress of matter, defined as $\pi_{ij} \doteq \delta_{im} \delta_{jn} \langle
v^n v^m \rangle - \delta_{ij} \langle v^2 \rangle / 3$.
Again, taking a double curl
would remove the vector and the linear scalar contributions. But since there are many nonlinear scalar terms which
survive, this does not seem promising.

The system of equations is closed once the equations of motion for
all sources of stress-energy are specified.
For CDM particles the evolution
of their phase space distribution follows from the geodesic equation of massive (non-relativistic) particles, given by
\be
\frac{\partial^2 x^i}{\partial \tau^2} + \mathcal{H} \frac{\partial x^i}{\partial \tau} + \delta^{ij} \bigg[\Psi_{,j}
- B'_j - \mathcal{H} B_j + 2 B_{\left[j,k\right]} \frac{\partial x^k}{\partial \tau}\bigg] = 0 \, , \label{eq:geodesic}
\ee
up to corrections beyond our approximation. Indices in square brackets are to be anti-symmetrized. In a pure
($\La$)CDM simulation, ${T_\mathrm{X}}_\mu^\nu = 0$, the tensor perturbations $h_{ij}$ do not feed back into the evolution of $T^\mu_\nu$
within our approximation.
Therefore, as long as one does not want to extract the tensor signal, the system of equations can be closed without
considering Eq.~(\ref{eq:h}). Of course, not all observables can then be reconstructed consistently, as $h_{ij}$ occurs
for instance in the null geodesic equation for light rays.

Interacting massive particles, like baryons, do not follow exact geodesics due to additional forces acting on them. These forces
have to be modeled according to the relevant microphysics. In this paper we do not concern ourselves with this problem which is
mainly relevant on small scales.

\section{Numerical implementation\label{sec:numerics}}

In this section we discuss in some more detail what it takes to convert a state-of-the-art Newtonian $N$-body code into
a general relativistic one in the sense of our approximation scheme, which is particularly tuned for cosmological applications.
$N$-body simulations have a long history alongside advancements in super computing technology, and it is beyond the scope of this
paper to give a comprehensive review of the subject. Some useful references are~\cite{Efstathiou:1985re}, \cite{Nbody},
and~\cite{Springel:2005nw,Springel:2008cc}. Further references can be found therein.

The $N$-body problem refers to the dynamics of $N$ particles which interact through long range forces. There are essentially two
ways to address the problem. One possibility is to consider the force acting on each particle as a sum of two-body forces, which
leads to the so-called ``particle-to-particle'' class of $N$-body algorithms. The alternative, referred to as the
``particle-to-mesh'' approach, is to construct a force field which permeates the entire volume of the simulation. To this end,
one projects the particle configuration onto a 3D grid (representing position space) and thus generates a discretized density field.
This grid information is used to compute the gravitational potential according to a discretized version of Eq.~(\ref{eq:poisson}).
The particles are then accelerated by the gradient of the potential, again suitably discretized and interpolated to the particle
positions. Based on these two approaches, many highly sophisticated algorithms have been developed. There also exist algorithms,
$P^3 M$ for example, which use a combination of both approaches.

In GR, gravitational interaction is mediated by the space-time metric. Therefore, it fits naturally into the framework of
the ``particle-to-mesh'' approach. Instead of constructing the Newtonian gravitational potential, however, we have to deal with a
multi-component field, the metric, which determines the gravitational acceleration according to Eq.~(\ref{eq:geodesic}).
In this paper, we discuss numerical solvers which allow to obtain the metric components on a structured mesh.
These can then be implemented in any suitable existing $N$-body code, which takes care of all remaining tasks, like
parallelization and mesh management.

From now on, we assume $\delta{T_\mathrm{X}}_\mu^\nu = 0$. That is, we develop a relativistic $N$-body algorithm for a
standard $\Lambda$CDM model. In this context, it is expected that Newtonian codes already give accurate results.
On small scales, there is a broad consensus on this. But on large scales, in particular scales which are outside the horizon when
the initial conditions are laid down, this is still under debate. Already within linear perturbation theory there are
relativistic corrections and the gauge dependence of the variables is
important on super horizon scales~\cite{Matsubara:2000pr,Chisari:2011iq,Bonvin:2011bg,Challinor:2011bk}.

Large-volume simulations will be important for the future large cosmological surveys, and so, not surprisingly, the horizon scale 
has already been crossed in recent Newtonian simulations. Examples are the Millennium
simulation~\cite{Springel:2005nw}, the Hubble Volume Project~\cite{Colberg:2000zv,Evrard:2001hu}, the Marenostrum Numerical
Cosmology Project~\cite{Gottlober:2006sx} and others~\cite{Park:2005ek}. It will be interesting to compare them with our
semi-relativistic approach and we think that this will provide a useful test of the validity of the Newtonian treatment
on cosmological scales. Furthermore, we hope to be able to give robust statements about the actual size of
relativistic corrections. In order to study extensions or alternatives to the standard model, the numerical scheme will have
to be amended for an appropriate treatment of $\delta{T_\mathrm{X}}_\mu^\nu \neq 0$.

Let us now briefly sketch the main components of our algorithm. At all times, the state of the system, representing
Cauchy data on a hypersurface of constant time which we may call a ``snapshot of the Universe'', is stored in two sets of data:
a particle list containing positions $x^i$ and peculiar velocities $v^i = \partial x^i / \partial \tau$ of $N$ test particles which are
used to sample the phase space distribution function\footnote{Unfortunately, representing the full distribution function of CDM in
six-dimensional (discretized) phase space is prohibitive, which is why the concept of test particles has to be introduced in the
first place.} of CDM, and a 3D grid (representing position space) holding values of all relevant fields at each grid point. A
standard particle-to-mesh projection allows to construct, from the particle list, the relevant moments of the CDM distribution
function on the grid, which we have defined as the ``bare'' quantities $\rho$, $u^i$, and $\pi_{ij}$. Some details on how this
construction works can be found in Appendix~\ref{app:equations}. 

Instead of merely solving for the Newtonian potential as in a
Newtonian particle-to-mesh code, we solve discretized (grid-based) versions of equations (\ref{eq:phi}), (\ref{eq:psi}), (\ref{eq:B}),
and possibly (\ref{eq:h}). The grid-based representation of the metric, suitably interpolated to the particle positions (see
Appendix~\ref{app:equations}), is then used to evolve the particle list by an infinitesimal time increment according
to the geodesic equation (\ref{eq:geodesic}), after which a next particle-to-mesh projection is performed in order to update the
metric, and so on.

In principle, our $N$-body scheme does not look very different from the Newtonian one. The equation for the particle
acceleration is slightly modified, and instead of an algorithm which solves the elliptic Poisson equation (\ref{eq:poisson}),
we need to implement algorithms which solve equations (\ref{eq:phi}), (\ref{eq:psi}), (\ref{eq:B}), and possibly (\ref{eq:h}).
With $\delta{T_\mathrm{X}}_\mu^\nu = 0$, equations (\ref{eq:phi}) and (\ref{eq:psi}) do not depend on $B_i$ or $h_{ij}$ directly,
so a natural ordering is to use them to obtain solutions for the scalar perturbations $\Phi$ and $\Psi$, which can later 
be plugged into the vector equation (\ref{eq:B}). Similarly, the vector perturbation $B_i$ can then be determined without any
knowledge of $h_{ij}$, and all the solutions can finally be used in Eq.~(\ref{eq:h}) to solve for the tensor perturbations.
Let us now discuss possible algorithmic solutions of these equations, one at a time.

\subsection{The scalar equations}

The two scalar equations (\ref{eq:phi}) and (\ref{eq:psi}) form a coupled system which is first order in time for $\Phi$, but
contains no time derivatives of $\Psi$. It is important to note that these equations are in fact a set of constraints,
since $\Phi$ and $\Psi$ themselves are not independent dynamical degrees of freedom. From their constrained dynamics we expect
that they evolve only very slowly, much slower even than the motion of individual $N$-body particles, for instance. It is therefore
sufficient, as a first approach, to construct a numerical update scheme which is only first order accurate in time. This approach
has the advantage that one can very easily decouple the update of $\Phi$ and $\Psi$: we will use Eq.~(\ref{eq:phi}) to update
$\Phi$, using the known $\Psi$ of the previous time step. The new $\Phi$ and $\Phi'$ obtained from this update step can then be
used in Eq.~(\ref{eq:psi}) to find a new solution for $\Psi$.

Under these premises, Eq.~(\ref{eq:phi}) is a nonlinear parabolic equation for $\Phi$ with a given source. If we forget about the
nonlinearity for a moment (which is anyway expected to remain very weak), it resembles a diffusion equation. This type
of mathematical problem is very well studied and can be solved with various standard methods found in the literature, see e.g.\
\cite{NumericalRecipes,Douglas1956,Douglas1962a}. 

The methods basically differ in the way how the differential operators are discretized, and the choice
can have a tremendous impact on the performance of the algorithm. For instance, a fully explicit scheme is unstable unless the
time step $d\tau$ satisfies a so-called Courant–Friedrichs–Lewy condition, which essentially states that information may not travel
by more than one lattice unit within one time step. For our equation, $\Phi$-information will basically propagate at the speed of
light, while on the other hand, particle velocities are roughly three orders of magnitude smaller. Therefore, it would take some
thousand iterations before a particle would have travelled only across a single cell of the lattice. This drawback can be overcome
by using implicit schemes, which can be shown to be unconditionally stable, hence allowing for much larger time steps. Since we
 violate the Courant–Friedrichs–Lewy condition of the explicit scheme, time resolution now becomes too coarse to follow the
time evolution of the small scales correctly, and different schemes show different behavior on these scales. We will choose a
scheme for which the small scales are driven towards their equilibrium solution ($\Phi' \rightarrow 0$), because this corresponds
to the behavior one expects from our discussion following Eq.~(\ref{eq:diffusiontime}).

Our choice is to use a slightly modified version of the alternating direction implicit (ADI) method described in
\cite{Douglas1956}.
There, an operator splitting method is used to construct an update rule which consists of three steps, i.e.\ one for each
space dimension. In each step, an implicit set of equations has to be solved along a different spatial direction. For a linear
parabolic operator, it can be shown that this algorithm is stable.
We modify this method slightly by adding the nonlinear terms, similar to how it is done in \cite{Douglas1962a}. Since the nonlinear
terms are expected to remain subdominant, this will not affect the stability of the algorithm. More details, including the discretized
version of Eq.~(\ref{eq:phi}), can be found in Appendix~\ref{app:equations}.

In order to perform a single update step from $\tau$ to $\tau + d\tau$ within the ADI framework, one has to invert a tridiagonal
matrix for each ``line'' of grid points along the implicit direction. This can be done efficiently using the Thomas algorithm,
which is easily parallelized for supercomputing applications \cite{MattorWH95}. In fact, the numerical solution of
Eq.~(\ref{eq:phi}) with this method is computationally not more expensive than solving the Poisson equation~(\ref{eq:poisson})
with the Fourier method. Since the ADI method operates in position space (rather than Fourier space), it is also compatible with
adaptive mesh refinement, which is a widely used approach to effectively increase the resolution of a simulation. We are therefore
confident to obtain a high performance of the algorithm within any state-of-the-art particle-to-mesh framework, competitive
with the standard Poisson solvers which are used in Newtonian codes.

Next we construct a new solution for $\Psi$
with the help of Eq.~(\ref{eq:psi}). Having already computed $\Phi$ and $\Phi'$, this is a nonlinear elliptic equation for
$\Psi$ with known coefficients and source. It can be solved again by standard methods, for instance using a multigrid algorithm
coupled to a nonlinear Gau\ss-Seidel relaxation scheme~\cite{NumericalRecipes}. Again, particular details can be found in Appendix~\ref{app:equations}. This approach should also compare well to the performance of a Newtonian code.

Having to solve two equations, (\ref{eq:phi}) and (\ref{eq:psi}), both requiring a numerical
effort comparable to the simple Poisson equation (\ref{eq:poisson}), the relativistic simulation will accordingly be
somewhat more expensive, but this is of course to be expected when we solve for two fields instead of one.

\subsection{The vector equation\label{sec:vec_imp}}

Let us now discuss how to obtain
a solution for the vector perturbation $B_i$, for given solutions $\Phi$ and $\Psi$. We use Eq.~(\ref{eq:B}), which is a linear elliptic equation for $B_i$.
Because of the product with the density perturbation on the right hand side, we use a method which operates in
position space, multigrid relaxation being an obvious possibility. We also have to avoid that numerical errors drive the
solution away from the transverse gauge. This can be done by subtracting the spurious longitudinal component from the
solution. It is given by the gradient of a scalar function $\chi$ which solves $\Delta \chi = \delta^{ij} \tilde{B}_{i,j}$,
where $\tilde{B}_i$ is a numerical solution with spurious longitudinal component. One then has $B_i = \tilde{B}_i - \chi_{,i}$.
All together, finding the solution $B_i$ amounts to solving four linear elliptic equations
(one for each component of $\tilde{B}_i$ and one for $\chi$) which are of the same numerical difficulty as Poisson's equation.

If one is not interested in the signal from tensor perturbations (gravitational waves), one can now close the loop
by implementing a particle update according to a suitable discretization of the geodesic equation (\ref{eq:geodesic}).
This is possible because $h_{ij}$ does not occur in the geodesic equation of massive particles at the level of our approximation.
Otherwise, one has to finally proceed with Eq.~(\ref{eq:h}). 

\subsection{The tensor equation}

Having solutions for $\Phi$, $\Psi$, and $B_i$ in hand, Eq.~(\ref{eq:h}) is a linear wave equation for the tensor modes $h_{ij}$
with a given source. Since it is linear, the different modes decouple in Fourier space where we are effectively dealing
with an ordinary differential equation (ODE) per Fourier-grid point. In addition, the gauge condition can be easily implemented in 
Fourier space. However, the hyperbolic nature of the equation means that in principle we would need a very high resolution
in time in order to resolve the rapid oscillations of the high-momentum waves that travel at the speed of light. 

This problem could be approached with the help of a standard stiff ODE solver, but we can gain more insight into the
behavior of the tensor modes by using a Green's function method:
In a $\La$CDM setup the source is given only by non-relativistic matter. Therefore,
the rapid oscillations come entirely from the homogeneous part of the equation while
the source term varies only slowly. Given the homogeneous solutions in Fourier space, $h_{\left(1\right)}(\tau)$ and
$h_{\left(2\right)}(\tau)$, one easily obtains the solution with source term by performing the integral
\be\label{eq:tens}
h(\tau) = \int_{\tau_{\rm in}}^\tau \frac{h_{\left(1\right)}(\tau)h_{\left(2\right)}(\tau') - h_{\left(1\right)}(\tau')h_{\left(2\right)}(\tau)}{W(\tau')}S(\tau')d\tau' \,,
\ee
where $W \doteq h_{\left(1\right)} h_{\left(2\right)}' - h_{\left(1\right)}' h_{\left(2\right)}$ is the Wronskian of the free
solutions $h_{\left(1\right)}$ and $h_{\left(2\right)}$, and $h(\tau)$ is the driven solution with vanishing initial values
at $\tau_\mathrm{in}$, which is what we need.
Here we have suppressed the tensor indices in $h$ and in the source but these are trivial in Fourier space.

In the matter dominated era, at redshift $z\gtrsim 2$, we can directly write down the homogeneous solutions. They
are simply given by
spherical Bessel functions, $h_{\left(1\right)} = (k\tau)^{-1}j_1(k\tau)$ and $h_{\left(2\right)} = (k\tau)^{-1}y_1(k\tau)$,
with Wronskian $W=k/(k\tau)^4$.

As is evident from the above expression, the oscillations are only driven when the frequency of the source term approximately
matches the one of the Green's function, which (for a non-relativistic matter source) happens only close to horizon crossing. At very
much slower variation of the source, the behavior of the system becomes that of a damped oscillator whose rest position
adiabatically follows a slowly varying external force. This means, as soon as the oscillations have died away, the displacement of
the oscillator is simply proportional to the instantaneous force.
In addition to this contribution which is sourced inside the horizon, we have free gravitational waves which have been produced
at horizon crossing, with energy density scaling as $\rho_h(k,\tau) \propto a^{-4}$.

\section{Plane-symmetric clustering\label{sec:results_variables}}

\begin{figure*}[tb]
\begin{minipage}{\textwidth}
\centerline{\includegraphics[width=0.95\columnwidth]{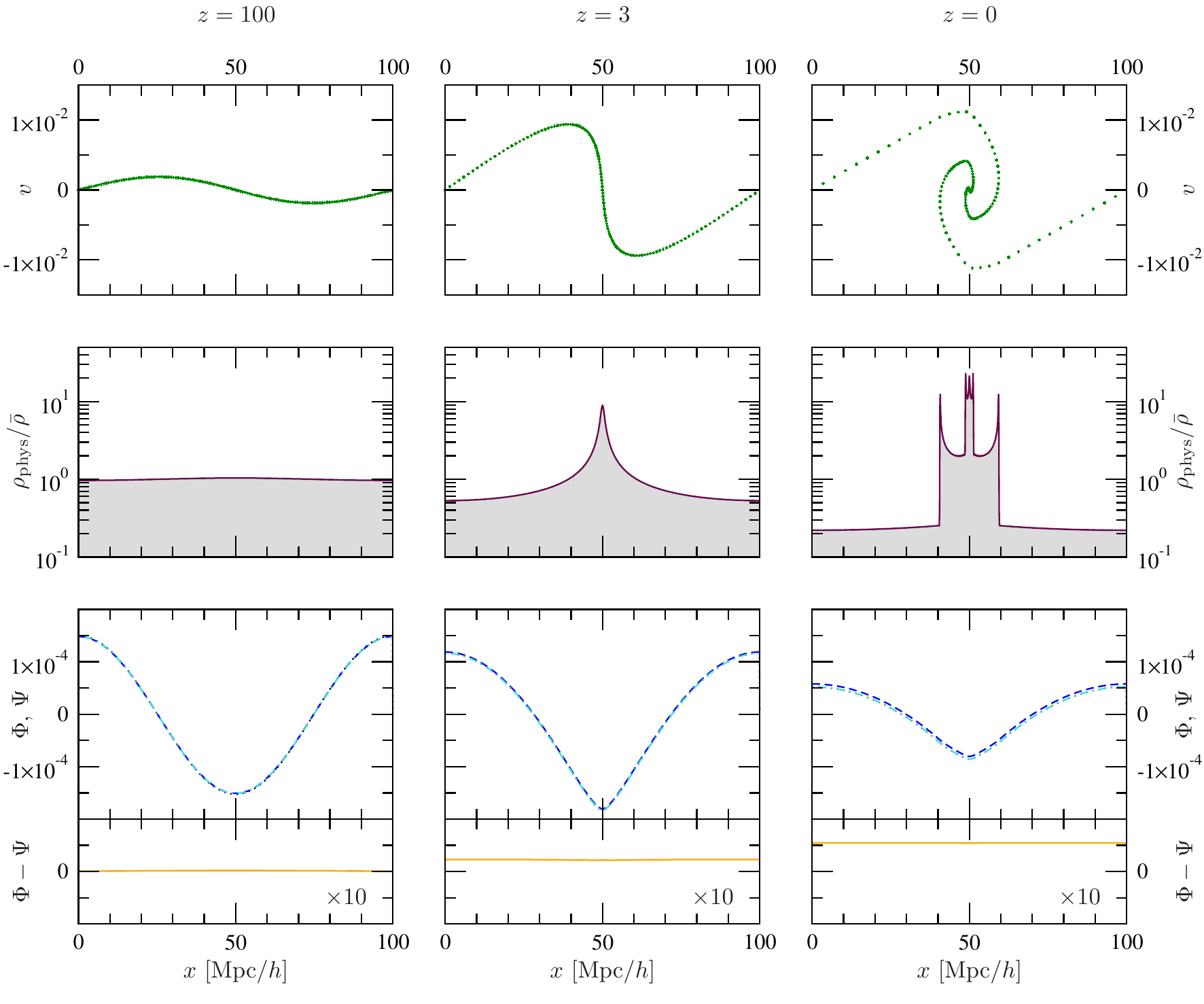}}
\end{minipage}
\caption{\label{fig:planewave} \small The plane wave example: the $x$-axis always refers to the non-trivial spatial direction
and is given in comoving units.
The three columns show snapshots at three different redshifts, $z=100$ (perturbations are still linear), $z=3$ (the density
perturbation is becoming nonlinear) and $z=0$ (today). The first row depicts the particle phase space of the simulation,
it is easy to see how shell crossing leads to a spiral-like structure in $(x,v)$. In the second row we plot the density.
As the central region around $x=50 \rm{Mpc}/h$ is overdense, this region undergoes gravitational collapse, resulting eventually
in shell crossing and the associated formation of caustics visible as sharp spikes at $z=0$. The bottom row shows the
behavior of the gravitational potentials $\Phi$ (dark blue dashed line) and $\Psi$ (light blue dot-dashed line).
We see that in this example the potentials always remain small, of the order of the initial value of $10^{-4}$. To display the
difference between the potentials more clearly, we plot it multiplied by a factor of 10 in the bottom-most panel
(using otherwise the same scale as for the potentials).
This difference mainly consists of a homogeneous mode that
can be interpreted as a correction to the background scale factor due to nonlinear effects.
The simulation was carried out with a resolution of $1024$ grid points and $16384$ particles (a representative subset is shown in
the phase space diagram).}
\end{figure*}

In order to test essential parts of our algorithm without needing to run many simulations on a supercomputer, we
restrict ourselves in this paper to the numerical simulation of plane-symmetric configurations. The planar symmetry trivializes
two of the three spatial dimensions, which reduces the computational requirements dramatically. However, it should be stressed
that imposing this symmetry is quite a strong restriction which precludes us from studying realistic models of cosmology.
Furthermore, the allowed configurations do not contain vector or tensor modes by construction. The results presented here
should therefore be understood as numerical tests of the algorithms, and can only be indicative of what will happen in more
realistic situations. We shall extend our investigation to the full three-dimensional case in forthcoming work.

As a first case, we consider an Einstein-de~Sitter background with an initial perturbation which is given by a single plane
wave with a comoving wavelength of $100 \mathrm{Mpc}/h$. 

We initialize the simulation at $z \simeq 5000$ using the linear solution
with an amplitude of $\Phi = \Psi = 1.5 \times 10^{-4}$, which is quite large in the sense that it will finally result in particle
velocities which reach $\sim 1\%$ of the speed of light. The generation of initial conditions in our general relativistic approach
has to be done in a slightly different way than for Newtonian simulations, in order to correctly account for gauge dependencies.
Details are given in Appendix~\ref{app:ics}.

Results are shown in Fig.~\ref{fig:planewave}. The fluctuation first grows linearly, but at $z=3$ has reached a nonlinear density
contrast of $\delta \sim 10$. Eventually, at $z=0$, two shell-crossings have occurred, which are highly nonlinear effects. As one
can see on the bottom panel, the two relativistic potentials $\Phi$ and $\Psi$ agree extremely well, up to the homogeneous mode in
$\Phi$ which reaches an amplitude of $\sim 5 \times 10^{-6}$ at $z=0$.

The amplitude of the homogeneous mode of $\Phi$ agrees well with our naive expectation that it should be governed by $v^2$. However,
its precise value could not have been computed simply by averaging $v^2$ from a Newtonian simulation and dressing Friedmann's
equations with a corresponding effective pressure and energy density. This is because other quadratic terms occur in the equations,
like $\delta^{ij} \Phi_{,i} \Phi_{,j}$, which are of the same order as $v^2$ and also contribute to the generation of
the homogeneous mode.

\begin{figure*}[tb]
\begin{minipage}{\textwidth}
\centerline{\includegraphics[width=0.95\columnwidth]{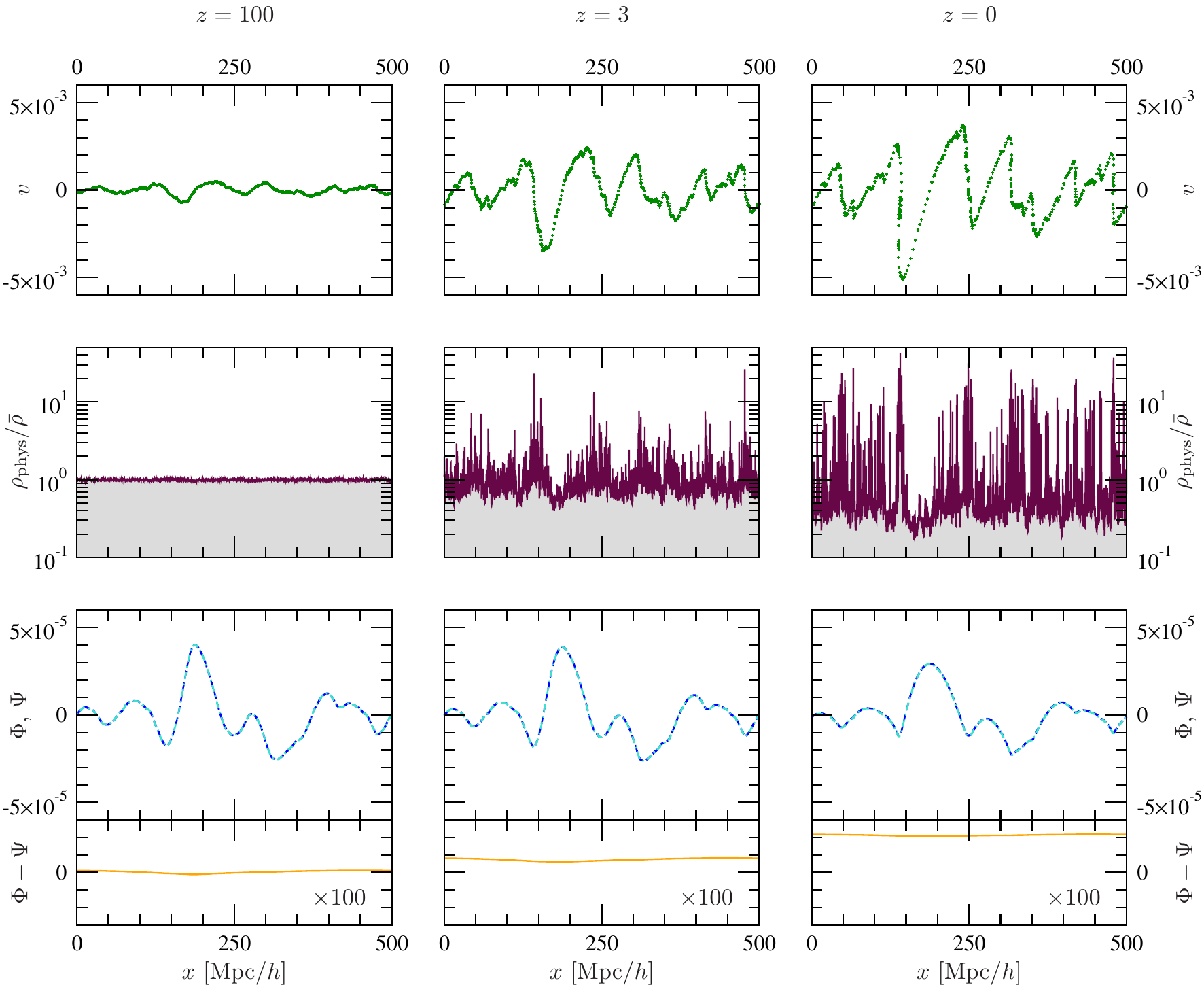}}
\end{minipage}
\caption{\label{fig:LCDM} \small A $\Lambda$CDM toy setup with planar symmetry: $\Omega_\Lambda \simeq 2/3$, and the initial power
spectrum for $\Phi$ is flat for $k < 0.075~h$/Mpc and scales like $k^{-4}$ for higher wavenumbers. As in Fig.\ \ref{fig:planewave}
we show the phase space (top row), the density (middle row) and the gravitational potentials (bottom row), for three different redshifts.
The last panel shows $\Phi - \Psi$  on the same scale as the potentials,
but multiplied by a factor of 100 (as indicated in the panel). The difference consists again mostly of a homogeneous
mode as in the plane wave example. The simulation was carried out with a resolution of $16384$ grid points and contained $131072$
particles.} 
\end{figure*}

\begin{figure*}[tb]
\begin{minipage}{\textwidth}
\centerline{\includegraphics[width=0.95\columnwidth]{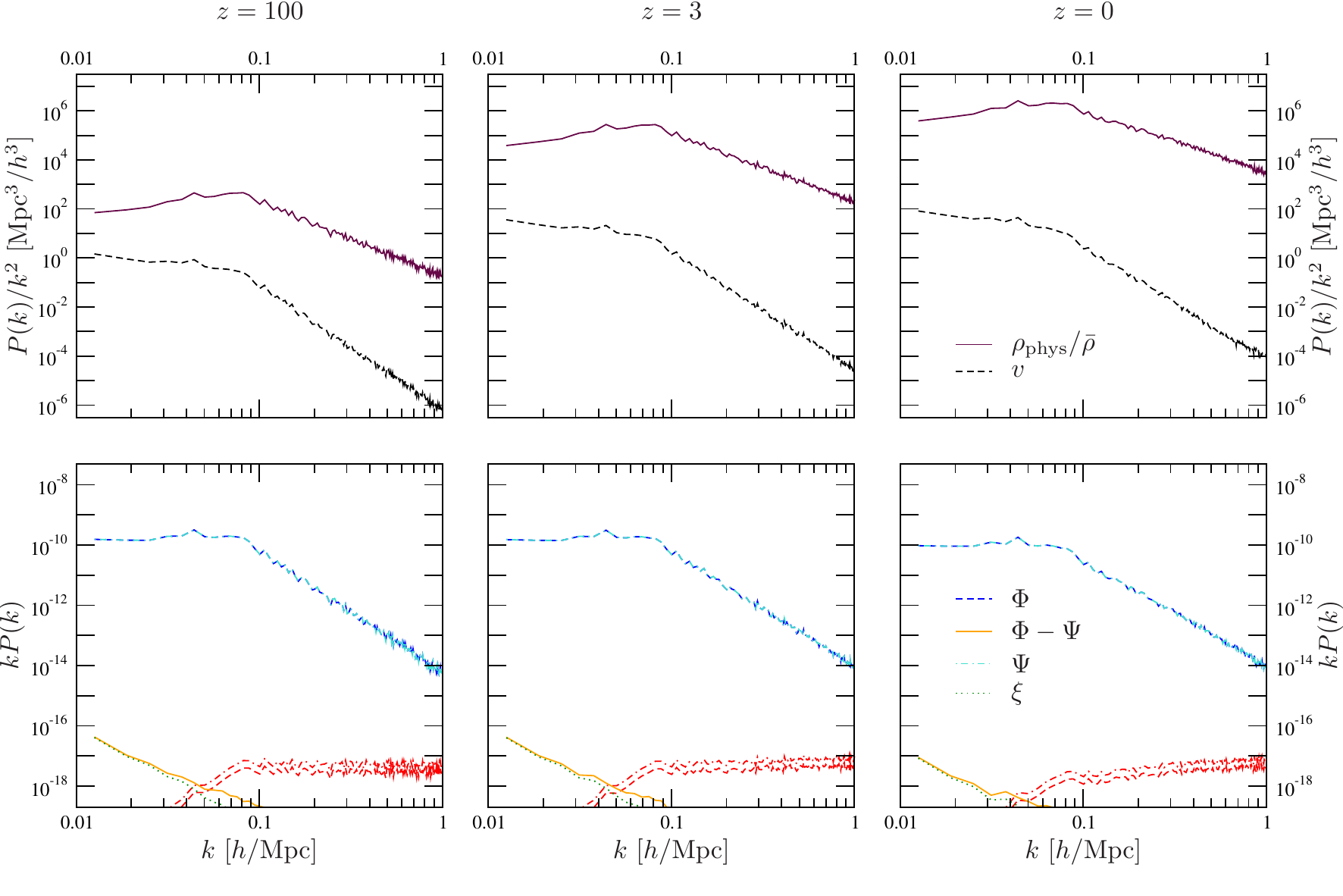}}
\end{minipage}
\caption{\label{fig:powerspectra} \small Power spectra averaged over 50 realizations of the $\Lambda$CDM toy setup for three
different redshifts.
The upper panels show the power spectra of the density (solid line) and of the velocity perturbations (dashed line). 
In the lower panels we plot the power spectra of the gravitational potentials $\Phi$ (dark blue dashed line) and $\Psi$
(light blue dot-dashed line) as well as their difference (orange solid line) compared to the second-order estimate
(green dotted line) based on $\xi$ from \cite{Green:2011wc}.
The red curves on the bottom of the plot show the truncation error (see text for definition) of the gravitational potentials
and indicate the expected numerical accuracy, and hence the level to which we can trust $\Phi-\Psi$.
Each simulation of the ensemble was carried out with a resolution of $4096$ grid points and contained $32768$ particles.}
\end{figure*}

As a second case, we study a plane-symmetric setup inspired by the $\La$CDM cosmology. Since a cosmological constant
is a homogeneous source of stress-energy, its effect enters the dynamics of cosmological perturbations only through a
modification of the background. We choose a $\La$CDM background
with $\Omega_\La = 1 - \Omega_m \simeq 2/3$ and draw a Gaussian random sample of initial plane wave perturbations which
are supposed to mimic a typical linear power spectrum of standard cosmology. However, as opposed to the standard case of an
isotropic power spectrum, our plane symmetry forces us to only include perturbations whose wave vector is perpendicular to the
plane of symmetry. Therefore, in order to obtain the right amplitude of perturbations at each scale, we choose
$k \langle | \Psi_k |^2 \rangle \propto k^3 P(k)$, where $P(k)$ denotes the usual isotropic linear power spectrum. Conversely,
when quoting results for the power spectra, we perform the angular integration over all wave vectors as if the perturbations
were statistically isotropic.

We choose an initial power spectrum where $k^3 P(k)$ is constant (scale invariant) at scales $k < 0.075 h/\mathrm{Mpc}$ and
decays as $k^{-4}$ on smaller scales. The simulation is initialized at $z=3900$ using linear theory, which in the real Universe
would be at the transition between radiation and matter domination. However, in our toy setup we do not include any radiation.
Linear theory guarantees that the linear solution will be in the growing mode (of the matter dominated solution) after a couple of
Hubble times, which means that radiation effects, for the purpose of CDM simulations, can be taken into account simply by
adjusting the amplitude of the growing mode. The physics of the radiation era and the transition to matter domination
are contained in the shape of the linear power spectrum.

Results are plotted in Figs.~\ref{fig:LCDM} and \ref{fig:powerspectra}. As in the previous example, we see that the difference
$\Phi-\Psi$ is dominated by the homogeneous mode of $\Phi$. The non-homogeneous ($k \neq 0$) component of $\Phi-\Psi$ has a very red spectrum.
Since we are in a setup where the ``dictionary'' of \cite{Green:2011wc}, given by their equations (2.40)--(2.42), should be valid,
we can actually estimate this component by computing a correction $\xi \doteq (\Psi-\Phi)_{k \neq 0}$ with the help of their equation (3.17),
where $\Delta^2\xi$ is given by a combination of quadratic terms in the velocity and in gradients of the Newtonian potential. To this end, we
first obtain the Newtonian quantities present in the latter equation by applying the ``dictionary'' in the reverse direction. As one can
see, the result obtained by this prescription agrees extremely well with our numerical result on the scales where it can be trusted.

The limited resolution introduces an intrinsic uncertainty in the numerical solutions for $\Phi$ and $\Psi$ on small scales. A benefit
of the multigrid scheme is that the size of this uncertainty can easily be estimated: we compare each solution at full resolution with
the one at half the resolution, interpolated to full resolution. The spectrum of the difference, the so-called truncation error,
is seen at the bottom of each plot in the lower panel of Fig.~\ref{fig:powerspectra} (for both $\Phi$ and $\Psi$). Certainly, the value
of the potentials should only be trusted up to the level indicated by this truncation error, and so should 
the value of $\Phi-\Psi$.
Because the difference between the potentials is so small, and has a red spectrum, it drops below the numerical accuracy on small scales.

\begin{figure}[tb]
 \includegraphics[width=\columnwidth]{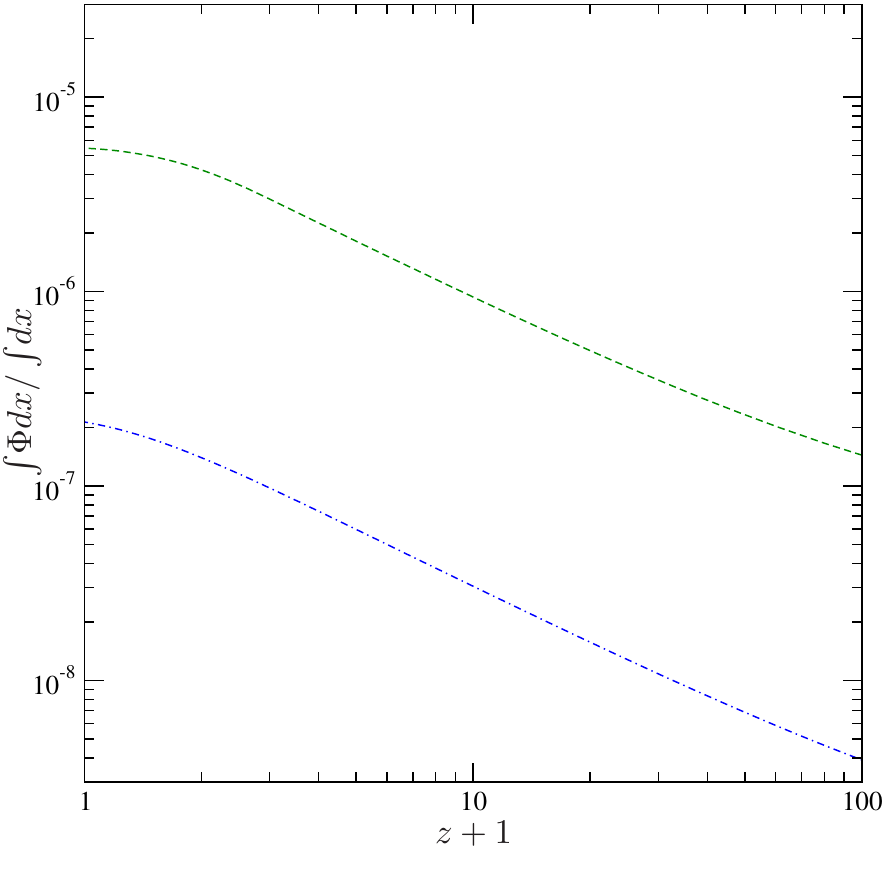}
\caption{\label{fig:zeromode} \small The homogeneous mode of $\Phi$ for the plane wave setup (green, dashed) and the $\Lambda$CDM setup
(blue, dot-dashed). This homogeneous mode shows the size of the ``backreaction'' effect for the plane-symmetric case, i.e.\ the
amount by which the average evolution of the perturbed universe differs from the exact FLRW solution.
More precisely, it can be absorbed into the background by a redefinition
$a \rightarrow a \left(1 + \int\! \Phi dx / \int\! dx\right)$.}
\end{figure}

As explained in Section~\ref{sec:scheme}, the homogeneous ($k = 0$) mode of $\Phi$ can be regarded as a correction to the
scale factor due to nonlinearities, and as such is a genuine backreaction effect.
Its amplitude is governed by $v^2$ and hence grows during linear evolution roughly $\propto a$, as can be seen
for our two numerical examples in Fig.~\ref{fig:zeromode}. However, as mentioned earlier, its value can only be
computed consistently by taking into account all contributions at a given order of approximation. At order $v^2$ there are,
for instance, also terms like $\delta^{ij} \Phi_{,i} \Phi_{,j}$ which are relevant.

\begin{figure}[tb]
 \includegraphics[width=\columnwidth]{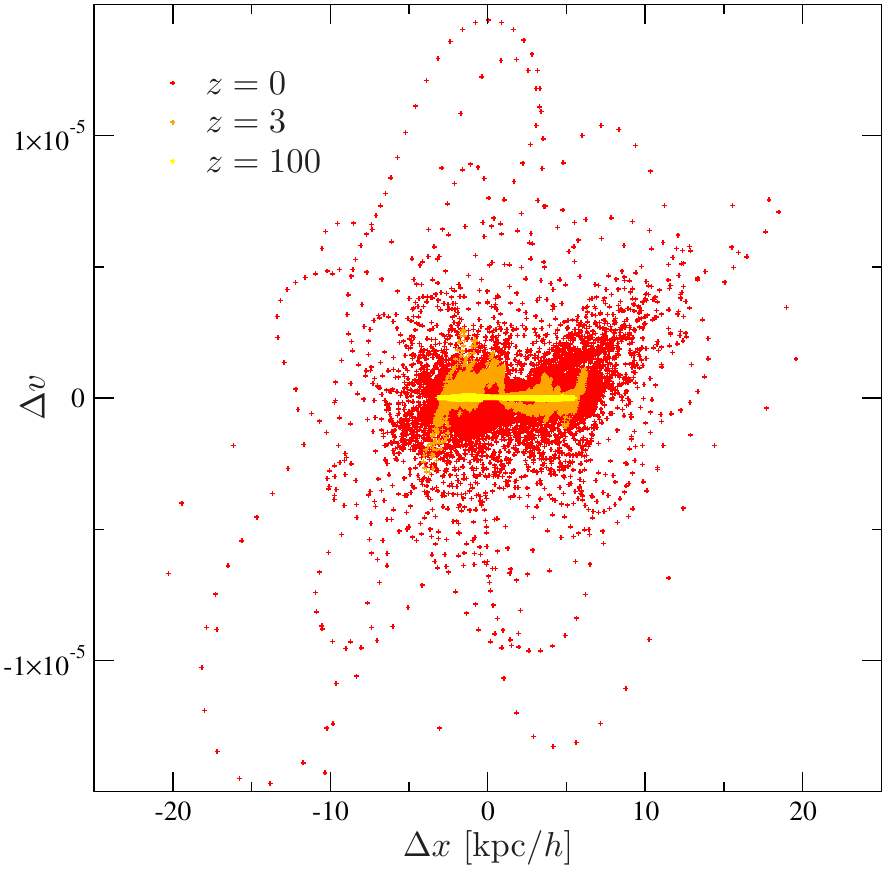}
\caption{\label{fig:pclcorr} \small Difference of the phase space coordinates for $N$-body particles in the relativistic
simulation of Fig.~\ref{fig:LCDM} and a purely Newtonian one, initialized on the same linear solution. The three colors yellow,
orange and red correspond to redshifts $z=100$, $z=3$ and $z=0$, respectively. As long as the solution is in the linear regime
(yellow, $z=100$), the difference in particle positions is essentially due to gauge dependence. At later times, however, particle
positions and velocities become affected by relativistic corrections. Nevertheless, the phase space distribution of particles
for the two different simulation techniques remains highly correlated, with velocities in agreement to well within a percent, and
particle positions within a few kpc/$h$.}
\end{figure}

In order to assess further the difference between the traditional Newtonian approach and our general relativistic one, we
also run a purely Newtonian simulation which is initialized on the identical linear solution as the example of Fig.~\ref{fig:LCDM}.
We then perform following comparison between the two simulations: on the initial data, before applying the initial infinitesimal
displacements, each $N$-body particle of the Newtonian simulation is paired with the corresponding $N$-body particle of the
relativistic simulation. Since the scalar velocity perturbation is gauge invariant, their initial velocities match perfectly;
on the other hand, as explained in Appendix~\ref{app:ics}, the initial particle displacement is gauge dependent and consequently
does not show perfect agreement. At $z=0$, however, the entire simulation box is well inside the particle horizon, and one therefore
expects gauge dependence to be weak. Preserving the initial pairing of particles, they are followed through the two separate
simulations. Fig.~\ref{fig:pclcorr} shows the separation of particle pairs in terms of phase space coordinates at $z=100$, $z=3$, and
$z=0$. Evidently, the $N$-body particles remain highly correlated between the two simulations. The rms separation in position space
rises from $\sim 2.5$ kpc/$h$ (comoving) in the initial data to $\sim 3.2$ kpc/$h$ at $z=0$, where the rms velocity difference
reaches $\sim 10^{-6}$.

This excellent agreement demonstrates that nonlinear effects of GR, at least when confined to the scalar sector, have an almost
negligible effect on the evolution of the $N$-body particle ensemble. On large scales where particle displacements can
be computed perturbatively, a similar level of agreement was found in \cite{Rigopoulos:2013nda}. Our results give a quantitative
indication that it extends to all scales on which gravitational fields remain weak, even if the particle distribution becomes highly
nonlinear. This underlines the sound performance of the Newtonian approximation. Note that the grid resolution of the two simulations
was $\sim 30$ kpc/$h$, which means that corresponding particles of the Newtonian and relativistic simulation are basically all found
within the same grid cell.

\section{Summary and outlook\label{sec:conclusions}}

In this paper we
develop and discuss the techniques for relativistic
$N$-body simulations in the weak field limit, keeping gravitational
perturbations to first order, their spatial gradients to second order and their second spatial derivatives to all orders.
This corresponds to keeping velocities to second order and the density contrast (which alone is expected to become large)
to all orders. In this way we are able to simulate structure formation in a General Relativistic context, taking along the
relativistic effects that can become relevant with current and future large cosmological surveys which will go out to
redshifts of $z \gtrsim 2$. Relativistic simulations are also a natural approach to include relativistic fields like those needed for
modified gravity simulations~\cite{Chan:2009ew,Li:2011vk,Lee:2012bm,Brax:2013mua,Puchwein:2013lza} or topological defects
\cite{Obradovic:2011mt}. In addition, these simulations allow to study whether backreaction effects become large and so to test
the backreaction scenario.

The numerical challenges in implementing our formalism as an extension of a standard $N$-body code are manageable,
but a full implementation still requires considerable effort. To test the scheme and to obtain initial results we have started
with an effectively one-dimensional implementation which allows to treat plane-symmetric situations. 
Because plane symmetry does not admit vector and tensor perturbations, and precludes us from studying
a realistic cosmological setup, we can only draw limited conclusions at this point. Our results show that, within this constrained
setup, relativistic corrections to the traditional Newtonian treatment remain very small. We highlight three different types of
corrections:

Firstly, fixing the background FLRW solution by assuming an exactly pressureless equation of state for CDM, the scalar potential
$\Phi$ dynamically acquires a homogeneous mode. The homogeneous mode quantifies the backreaction and shows how the evolution of
the averaged ``background'' slightly differs from the reference FLRW solution.
With a relative amplitude of $\sim 10^{-7}$ in the $\La$CDM example, which is commensurate to the expected order of magnitude $\sim v^2$,
it remains observationally irrelevant, but we want to make the point that its exact value can only be calculated consistently by taking
into account all relevant terms at a given order. At order $v^2$ there are, for instance, terms like $\delta^{ij}\Phi_{,i}\Phi_{,j}$
which are important.

A second sign of relativistic corrections comes from considering the inhomogeneous component of $\Phi-\Psi$. As is well known, in linear perturbation theory $\Phi-\Psi$ is sourced by anisotropic
stress, which vanishes for a CDM source at linear order. It is therefore generated only at the nonlinear level (e.g. \cite{Ballesteros:2011cm}). However, as explained
in \cite{Green:2011wc}, it can be accurately estimated from Newtonian quantities. A comparison of our numerical results with this
estimate is an excellent test for our algorithms. We demonstrate that the numerical scheme is able to give highly accurate
results for this second order quantity, down to the truncation error which is introduced by discretization.

Finally, as a third way to quantify the difference between relativistic and purely Newtonian simulations, we look at the pairwise
phase space correlation of individual particles between both simulation types when initialized on the same linear solution. We find that
relativistic effects on the phase space trajectory of particles remain completely tolerable. In fact, the
coordinates of individual $N$-body particles remain correlated to within a few kpc$/h$ (comoving), and their velocities to
within $1\%$.

All these observations have to be considered under the premise that our plane-symmetric setups are limited to scalar
perturbations. However, recent quantitative analyses strongly suggest \cite{Bruni:2013mua} that the inclusion of vector modes will
also only have a small effect on the $N$-body dynamics, although there may be some potential for observing $B_i$ through its effect
on photon propagation. The same holds true for the tensor perturbations.
A detailed quantitative comparison with 3-dimensional Newtonian $N$-body simulations is in preparation.
There we shall also calculate the induced vector and tensor perturbation spectra.

While observing relativistic effects within the context of $\La$CDM standard cosmology will be challenging,
we should keep in mind that relativistic simulations of structure formation have a great potential in testing possible extensions
to the standard model. For instance, cosmic neutrinos or a warm component of Dark Matter may constitute a semi-relativistic source
which is relevant during the process of structure formation. In the sector of Dark Energy it is also very important to be able
to test various alternatives to a cosmological constant, and again it seems most promising to use large scale structure as a
probe. Once the numerical scheme for General Relativistic simulations is fully implemented, the challenge may lie in the modelling
and evolution of the stress-energy tensor of the additional sources one wants to consider.

\begin{acknowledgments}
It is a pleasure to thank Chris Clarkson and Roy Maartens for interesting comments.
J.A.\ acknowledges funding from the German Research Foundation (DFG) through the research fellowship \mbox{AD~439/1-1}. This
work is supported  by the Swiss National Science Foundation.
\end{acknowledgments}

\appendix

\section{Lattice equations\label{app:equations}}

In this appendix, we explicitly present the finite-difference operations which define the various algorithms presented in this paper.
We assume that the code represents continuous fields (like, e.g., the metric) on a structured mesh of rank three, which can basically
be identified with a comoving coordinate grid covering a spacelike hypersurface. We make use of the convenient notation
$f^\mathbf{n}_\mathbf{i,j,k} \doteq f(\tau_\mathbf{n}, x^i_\mathbf{i,j,k})$, where $\mathbf{n}$ is a discrete index labelling the time steps,
while $\mathbf{i}, \mathbf{j}, \mathbf{k}$ are discrete indices labelling the grid points of the mesh. In some cases, a quantity may be defined
with a fractional index whose meaning should be clear from the context. The grid spacing is given in units of $dx^1$, $dx^2$, $dx^3$, and
the size of a time step is denoted as $d\tau$.

The evolution of dark matter will be given by the notion of test particles which move on time-like geodesics. Their
coordinates $x^i$ and velocities $v^i = \partial x^i / \partial \tau$ are stored in a list and can take continuous values. Like in a standard
particle-to-mesh $N$-body framework, the code therefore has to provide appropriate projection and interpolation operators in order to
connect between grid-based and particle-based information. If dark matter is represented in
a different way, or if one wants to include baryons or other types of interacting matter, it should still be more or less straightforward
to implement the correct gravitational acceleration. We leave it as an exercise to the reader to figure out the appropriate prescription
for their favorite hydrodynamical scheme.

\subsection{ADI scheme}

The Alternate-Direction-Implicit (ADI) update for $\Phi$ is carried out by splitting the operation into three steps, one step
for each spatial dimension. The idea is that each step amounts to only solving a linear tridiagonal system, which can be done very
efficiently using the Thomas algorithm. There is no unique way of performing the split. We simply follow \cite{Douglas1956} and add
the nonlinear terms in the most straightforward fashion. There may be better splitting procedures, and it may be worthwhile to
investigate in this direction. With three spatial dimensions, the update makes use of two intermediate solutions,
$\Phi^\mathbf{n+\frac{1}{3}}$ and $\Phi^\mathbf{n+\frac{2}{3}}$. These should not be interpreted as solutions at fractional time
steps, but rather as auxiliary surrogates for $\Phi^\mathbf{n+1}$. They obey following finite difference equations:
\begin{widetext}
\begin{subequations}
 \begin{multline}
 \left(1 + 4 \Phi^\mathbf{n}_\mathbf{i,j,k}\right) \left(\frac{\Phi^\mathbf{n+\frac{1}{3}}_\mathbf{i-1,j,k}
+ \Phi^\mathbf{n+\frac{1}{3}}_\mathbf{i+1,j,k} - 2 \Phi^\mathbf{n+\frac{1}{3}}_\mathbf{i,j,k}}{\left(dx^1\right)^2}
+ \frac{\Phi^\mathbf{n}_\mathbf{i,j-1,k} + \Phi^\mathbf{n}_\mathbf{i,j+1,k}
- 2 \Phi^\mathbf{n}_\mathbf{i,j,k}}{\left(dx^2\right)^2} + \frac{\Phi^\mathbf{n}_\mathbf{i,j,k-1} + \Phi^\mathbf{n}_\mathbf{i,j,k+1}
- 2 \Phi^\mathbf{n}_\mathbf{i,j,k}}{\left(dx^3\right)^2}\right) \\ - 3 \mathcal{H} \frac{\Phi^\mathbf{n+\frac{1}{3}}_\mathbf{i,j,k} -
\Phi^\mathbf{n}_\mathbf{i,j,k}}{d\tau} - 3 \mathcal{H}^2 \Psi^\mathbf{n}_\mathbf{i,j,k} +
\frac{3}{2} \left(\frac{\left(\Phi^\mathbf{n}_\mathbf{i+1,j,k} - \Phi^\mathbf{n}_\mathbf{i-1,j,k}\right)^2}{4 \left(dx^1\right)^2}
+ \frac{\left(\Phi^\mathbf{n}_\mathbf{i,j+1,k} - \Phi^\mathbf{n}_\mathbf{i,j-1,k}\right)^2}{4 \left(dx^2\right)^2}
+ \frac{\left(\Phi^\mathbf{n}_\mathbf{i,j,k+1} - \Phi^\mathbf{n}_\mathbf{i,j,k-1}\right)^2}{4 \left(dx^3\right)^2}\right) \\ = 4 \pi G a^2 \bar{\rho}
\left[\delta^\mathbf{n}_\mathbf{i,j,k} + 3 \Phi^\mathbf{n}_\mathbf{i,j,k} \left(1 + \delta^\mathbf{n}_\mathbf{i,j,k}\right) +
\frac{1}{2} \left(1 + \delta^\mathbf{n+\frac{1}{2}}_\mathbf{i,j,k}\right)
\langle v^2 \rangle^\mathbf{n+\frac{1}{2}}_\mathbf{i,j,k}\right] \, ,\label{eq:phi-discrete}
\end{multline}
\be
 \left(1 + 4 \Phi^\mathbf{n}_\mathbf{i,j,k}\right) \frac{\Phi^\mathbf{n+\frac{2}{3}}_\mathbf{i,j-1,k}
+ \Phi^\mathbf{n+\frac{2}{3}}_\mathbf{i,j+1,k} - 2 \Phi^\mathbf{n+\frac{2}{3}}_\mathbf{i,j,k}}{\left(dx^2\right)^2} =
\left(1 + 4 \Phi^\mathbf{n}_\mathbf{i,j,k}\right) \frac{\Phi^\mathbf{n}_\mathbf{i,j-1,k} + \Phi^\mathbf{n}_\mathbf{i,j+1,k}
- 2 \Phi^\mathbf{n}_\mathbf{i,j,k}}{\left(dx^2\right)^2} + 3 \mathcal{H} \frac{\Phi^\mathbf{n+\frac{2}{3}}_\mathbf{i,j,k} -
\Phi^\mathbf{n+\frac{1}{3}}_\mathbf{i,j,k}}{d\tau}\, ,
\ee
\be
 \left(1 + 4 \Phi^\mathbf{n}_\mathbf{i,j,k}\right) \frac{\Phi^\mathbf{n+1}_\mathbf{i,j,k-1}
+ \Phi^\mathbf{n+1}_\mathbf{i,j,k+1} - 2 \Phi^\mathbf{n+1}_\mathbf{i,j,k}}{\left(dx^3\right)^2} =
\left(1 + 4 \Phi^\mathbf{n}_\mathbf{i,j,k}\right) \frac{\Phi^\mathbf{n}_\mathbf{i,j,k-1} + \Phi^\mathbf{n}_\mathbf{i,j,k+1}
- 2 \Phi^\mathbf{n}_\mathbf{i,j,k}}{\left(dx^3\right)^2} + 3 \mathcal{H} \frac{\Phi^\mathbf{n+1}_\mathbf{i,j,k} -
\Phi^\mathbf{n+\frac{2}{3}}_\mathbf{i,j,k}}{d\tau}\, .
\ee
\end{subequations}
Indeed, each equation is a linear tridiagonal problem. The stability of the scheme for a linear parabolic operator has been proven in
\cite{Douglas1956}. This property can in principle be lost when nonlinear terms are present in the equation. However, we expect that in our
case the nonlinearities remain sufficiently subdominant that this is not an issue.

Note that in the plane symmetric setup, the last two equations become trivial and simply imply
$\Phi^\mathbf{n+\frac{1}{3}} = \Phi^\mathbf{n+\frac{2}{3}} = \Phi^\mathbf{n+1}$, i.e.\ the auxiliary solutions are
redundant.

\subsection{Multigrid}

In order to solve Eq.~(\ref{eq:psi}), we follow \cite{NumericalRecipes} and use a nonlinear multigrid scheme (``FAS algorithm'')
coupled to a Newton-Gau\ss-Seidel relaxation method. The key equation is (20.6.43) of \cite{NumericalRecipes}, identifying
$u_\mathbf{i}$ with our sought-after $\Psi^\mathbf{n}_\mathbf{i,j,k}$ and using the discretized operator
 \begin{multline}
 \mathcal{L}(\Psi^\mathbf{n}_\mathbf{i,j,k}) \doteq \left(1 + 2 \Phi^\mathbf{n}_\mathbf{i,j,k} - 2 \Psi^\mathbf{n}_\mathbf{i,j,k}\right)
 \left(\frac{\Psi^\mathbf{n}_\mathbf{i-1,j,k} + \Psi^\mathbf{n}_\mathbf{i+1,j,k} - 2 \Psi^\mathbf{n}_\mathbf{i,j,k}}{\left(dx^1\right)^2}
 + \ldots\right)\\
- \frac{\left(\Psi^\mathbf{n}_\mathbf{i+1,j,k} - \Psi^\mathbf{n}_\mathbf{i-1,j,k}\right) \left(\Phi^\mathbf{n}_\mathbf{i+1,j,k}
- \Phi^\mathbf{n}_\mathbf{i-1,j,k} + \Psi^\mathbf{n}_\mathbf{i+1,j,k} - \Psi^\mathbf{n}_\mathbf{i-1,j,k}\right)}{4 \left(dx^1\right)^2} - \ldots\\
+ \frac{1}{\mathcal{H}} \Bigg[\left(1 + 4 \Phi^\mathbf{n}_\mathbf{i,j,k}\right)
\left(\frac{\Phi^\mathbf{n}_\mathbf{i-1,j,k} - \Phi^\mathbf{n-1}_\mathbf{i-1,j,k} + \Phi^\mathbf{n}_\mathbf{i+1,j,k}
- \Phi^\mathbf{n-1}_\mathbf{i+1,j,k} - 2 \Phi^\mathbf{n}_\mathbf{i,j,k} + 2 \Phi^\mathbf{n-1}_\mathbf{i,j,k}}{d\tau \left(dx^1\right)^2} + 
\ldots\right) \\+ 4 \frac{\Phi^\mathbf{n}_\mathbf{i,j,k} - \Phi^\mathbf{n-1}_\mathbf{i,j,k}}{d\tau} \left(
\frac{\Phi^\mathbf{n}_\mathbf{i-1,j,k} + \Phi^\mathbf{n}_\mathbf{i+1,j,k} - 2 \Phi^\mathbf{n}_\mathbf{i,j,k}}{\left(dx^1\right)^2} +
\ldots\right) \\ + 3\frac{\left(\Phi^\mathbf{n}_\mathbf{i+1,j,k} - \Phi^\mathbf{n-1}_\mathbf{i+1,j,k} - \Phi^\mathbf{n}_\mathbf{i-1,j,k}
+ \Phi^\mathbf{n-1}_\mathbf{i-1,j,k}\right)\left(\Phi^\mathbf{n}_\mathbf{i+1,j,k} - \Phi^\mathbf{n}_\mathbf{i-1,j,k}\right)}{4 d\tau \left(dx^1\right)^2} + \ldots\Bigg] \\
+ \frac{4 \pi G a^2 \bar{\rho}}{\mathcal{H}} \Bigg[\left(1 + 3 \Phi^\mathbf{n}_\mathbf{i,j,k}\right)
\left(\frac{\left(\left(1 + \delta\right) \langle \gamma v^1 \rangle\right)^\mathbf{n-\frac{1}{2}}_\mathbf{i+\frac{1}{2},j,k} - \left(\left(1 + \delta\right) \langle \gamma v^1 \rangle\right)^\mathbf{n-\frac{1}{2}}_\mathbf{i-\frac{1}{2},j,k}}{dx^1} + \ldots\right)\\
+ \frac{\left(\left(1 + \delta\right) \langle \gamma v^1 \rangle\right)^\mathbf{n-\frac{1}{2}}_\mathbf{i+\frac{1}{2},j,k} + \left(\left(1 + \delta\right) \langle \gamma v^1 \rangle\right)^\mathbf{n-\frac{1}{2}}_\mathbf{i-\frac{1}{2},j,k}}{2}
\times \frac{\Psi^\mathbf{n}_\mathbf{i+1,j,k} - \Psi^\mathbf{n}_\mathbf{i-1,j,k}}{2 dx^1} + \ldots - 3 \delta^\mathbf{n}_\mathbf{i,j,k}
\frac{\Phi^\mathbf{n}_\mathbf{i,j,k} - \Phi^\mathbf{n-1}_\mathbf{i,j,k}}{d\tau}\Bigg] \, . \label{eq:discrete-psi}
 \end{multline}
\end{widetext}
Here, the ellipsis indicates that the previous term has to be written also for the other two directions $x^2$ and $x^3$ accordingly.
Note that our operator $\mathcal{L}$ is only accurate to first order in time although one could easily modify it such that it would be
accurate to second order. However, since the ADI algorithm used for the evolution of $\Phi$ is a first order scheme, the implementation of
a second order scheme for $\Psi$ would probably not increase the overall accuracy.

We perform the relaxation sweep in the ``checkerboard'' fashion, such that updated values on neighboring sites are already available
when updating site $\left(\mathbf{i,j,k}\right)$. We also store the approximate solution for all multigrid levels between time steps,
such that the previous solution can be used as an initial guess in the next step. Since the potential $\Psi$ varies slowly in time, this
initial guess is already very close to the final solution. Tracking the solution in such a way, we found that the multigrid algorithm
generally needs no more than a single V-cycle to meet the convergence criterion again after one time step. Note that the additional
memory to store the coarse-grid approximations, in 3D, is bounded by $1/7 \simeq 15\%$ of the memory consumed by the full resolution.

\subsection{Loop structure}
As usual, we perform the particle updates in a leap-frog fashion, i.e.\ we associate particle positions and acceleration to integer
time steps, while velocities are associated to half-integer time steps. A complete loop of one time step of our algorithm can be
sketched as follows:
\begin{itemize}
 \item update velocities:
 \be
   \left(v^i\right)^\mathbf{n+\frac{1}{2}} = \frac{\left(v^i\right)^\mathbf{n-\frac{1}{2}}
   \left(1 - \frac{d\tau}{2}\mathcal{H}\right) - d\tau \nabla \Psi^\mathbf{n}}{1 + \frac{d\tau}{2}\mathcal{H}}
 \ee
 \item update positions by half a step:
 \be
   \left(x^i\right)^\mathbf{n+\frac{1}{2}} = \left(x^i\right)^\mathbf{n} + \frac{d\tau}{2} \left(v^i\right)^\mathbf{n+\frac{1}{2}}
 \ee
 \item do particle-to-mesh projection for $\left(\left(1 + \delta\right) \langle \gamma v^i\rangle\right)^\mathbf{n + \frac{1}{2}}$ (face-centered, i.e.\ onto a
 mesh shifted by half a grid unit in direction $x^i$) and $\left(1 + \delta^\mathbf{n + \frac{1}{2}}\right) \langle v^2 \rangle^\mathbf{n + \frac{1}{2}}$
 (cell-centered)
 \item update $\Phi^\mathbf{n} \rightarrow \Phi^\mathbf{n+1}$
 \item update positions by half a step:
 \be
   \left(x^i\right)^\mathbf{n+1} = \left(x^i\right)^\mathbf{n+\frac{1}{2}} + \frac{d\tau}{2} \left(v^i\right)^\mathbf{n+\frac{1}{2}}
 \ee
 \item do particle-to-mesh projection for $\delta^\mathbf{n+1}$
 \item compute $\Psi^\mathbf{n+1}$
 \item $\mathbf{n++}$ \ldots
\end{itemize}
The gradient of $\Psi$, and other metric terms, have to be interpolated to the particle positions when updating the velocities,
as will be specified in the next subsection. The loop shown here only includes the scalar degrees of freedom and is sufficient for the
plane-symmetric case. When vector modes are taken into account, the update of the particle velocities has to be modified accordingly,
and the vector component of the metric has to be computed at an appropriate instance within the loop. Finally, if desired, one can also
insert the evolution step of the tensor component.

\subsection{Particle-to-mesh projection and force interpolation\label{app:p2m}}

In order to establish the connection between grid-based and particle-based information, we make use of some standard approaches
which are detailed in \cite{HockneyEastwood}. There, a systematic hierarchy of prescriptions which assign particle properties to
grid points is developed, explicitly discussing nearest-grid-point (NGP), cloud-in-cell (CIC) and triangular-shaped-particle (TSP)
as the first three members of this hierarchy. Generally speaking, one can trade a reduction of discretization errors for a more
complicated assignment scheme.

As a guiding principle, since we are explicitly using conservation of stress-energy to arrive at Eq.~(\ref{eq:psi}), we aim for
an assignment scheme which guarantees to satisfy the discrete version of the continuity equation,
\be
\left[a^3 \rho\right]' + a^3 \left[\rho u^i\right]_{,i} = 0 \, ,
\ee
where the divergence of the momentum density is discretized precisely in the same way as in Eq.~(\ref{eq:discrete-psi}).

Choosing to construct $a^3 \rho$ with the TSP assignment scheme, one can analytically check that above equation holds identically
if we construct $a^3 \rho u^i$ using CIC assignment in direction $x^i$ (on the face-centered grid), and TSP assignment in the remaining
directions. We trivially add the $v^2$-corrections by multiplying the result for each particle by $1 + v^2/2$, which accounts for the
kinetic energy density or Lorentz factor, respectively.

The interpolation of the grid-based field information to the particle positions must be done in such a way that the resulting
force does not include a relevant self-force. According to \cite{HockneyEastwood}, it is generally sufficient if the force
interpolation scheme is at most of the same polynomial order as the particle-to-mesh assignment scheme, at least when all equations
are linear. We do not expect that nonlinearities introduce an instability in our case, and so far our numerical simulations have shown
no indication for such an instability. In practice, we used CIC force interpolation.

\section{Initial conditions for Newtonian and relativistic simulations\label{app:ics}}

The smallness of perturbations in the early Universe allow us to pose the initial conditions at a time when perturbation
theory is valid. Traditionally, the well-known solutions of linear perturbation theory have been used for this purpose, although
people have also started to consider results from second order perturbation theory \cite{Crocce:2006ve}. We shall discuss only the
former approach, although ultimately, implementing the latter one is certainly desirable as it would guarantee that all second order
terms within our framework are accurate. However, for the purpose of this work, we simply always chose the initial redshift high
enough that second order terms can safely be neglected, giving our simulation enough time to evolve to the nonlinear solution on
its own.

As shown by Bardeen \cite{Bardeen:1980kt}, linear cosmological perturbations can be characterized completely in terms of
gauge-invariant quantities. Once the linear solutions for these quantities have been determined, it is just a matter of relating
gauge-dependent quantities to these solutions in order to obtain the linear solutions in any gauge. In the longitudinal gauge
which we use in our relativistic framework, these relations are given by
\begin{subequations}
\label{eq:Bardeen}
 \bea
  \Psi &=& \Psi Q^{\left(S\right)} \, ,\\
  \Phi &=& \Phi  Q^{\left(S\right)}\, ,\\
  B_i &=& \sigma^{(V)} Q_i^{\left(V\right)} \, ,\\
  h_{ij} &=& 2 H^{\left(T\right)} Q_{ij}^{\left(T\right)} \, ,\\
  u^i &=& V Q^{\left(S\right) i} + V^{\left(V\right)} Q^{\left(V\right) i} \, , \\
  \delta + 3 \Phi &=& D_g Q^{\left(S\right)}\, ,
 \eea
\end{subequations}
where the functions $Q^{\left(S\right)}$, $Q_i^{\left(V\right)}$, $Q_{ij}^{\left(T\right)}$ denote the scalar, vector and
tensor Fourier modes, respectively, and the amplitudes are given in the notation of \cite{2008cmbg.book}.
In particular, we use the same notation for $\Phi$, $\Psi$ and its Fourier transform.
$V$ and $V^{\left(V\right)}$ denote the gauge-invariant amplitudes of the scalar (curl-free) and vector
(divergence-free) parts of a Helmholtz decomposition of the velocity field, respectively, and $D_g$ is a possible gauge-invariant
amplitude of the density perturbation. The left hand side of the last line is
obtained by truncating our expression for the energy density, Eq.~(\ref{eq:T00}), at linear order. At this level, the anisotropic stress
of non-relativistic matter is negligible, as is its effective pressure.

It is usually assumed that vector modes are not significantly excited in the early Universe or have had time to decay until
the onset of matter domination. One can then consistently set $\sigma^{(V)} = V^{\left(V\right)} = 0$ during linear evolution,
which sets $B_i = 0$ and implies that $u^i$ is given by the gradient of a scalar function.

At linear order, scalar, vector and tensor equations decouple and can be analyzed separately. Furthermore, in Fourier space,
Einstein's equations reduce to a system of ordinary differential equations for the gauge-invariant amplitudes in each
sector. In the scalar sector, this system is second order in time, giving two independent solutions. For the matter dominated era,
these can be found analytically,
\begin{subequations}
\label{eq:linearsolutions}
 \bea
\Phi &=& \Psi = c_1 + c_2 \tau^{-5} \, ,\\
 V&=& \frac{k}{3}c_1\tau - \frac{k}{2}c_2 \tau^{-4} \, ,\\
 D_g &=& -\frac{k^2}{6} \left(1 + 3 \frac{\mathcal{H}^2}{k^2}\right) c_1 \tau^2
 - \frac{k^2}{6} \left(1 - \frac{9 \mathcal{H}^2}{2 k^2}\right) c_2 \tau^{-3} \, , \nonumber\\
 \eea
\end{subequations}
where $c_1$ and $c_2$ are the two constants of integration. The solution proportional to $c_2$ decays rapidly and is usually
discarded. One then arrives at the well-known result that the gauge-invariant Bardeen potentials $\Psi$ and $\Phi$
are constant (and equal) during matter domination. The linear solution for the other gauge-invariant quantities is uniquely
determined once these potentials are specified.

The vector equations are only first order in time, and the solution for the ``frame-dragging potential'' reads
$\sigma^{(V)} \propto \tau^{-4}$ in the matter dominated era. Since this solution decays, $B_i = 0$ as initial condition
remains a reasonable choice.

The tensor equations are again second order in time. As already mentioned in Section~\ref{sec:numerics}, the two independent
free solutions for $H^{\left(T\right)}$ are given by $(k\tau)^{-1}j_1(k\tau)$ and $(k\tau)^{-1}y_1(k\tau)$, with $j_1$ and
$y_1$ spherical Bessel functions. Both solutions oscillate and decay inside the horizon, $k \tau \gg 1$. However, outside the
horizon, only the latter one decays, while the former one remains approximately constant. Its amplitude is determined by early
Universe physics. Usually, one assumes that these super-horizon modes are generated during inflation, at the time when they exit
the horizon. They can be constrained, e.g., by observations of the cosmic microwave background. Thus far, only upper limits have
been obtained, and observations are therefore compatible with the choice $h_{ij} = 0$ as initial condition on all scales.

Our strategy to generate initial data is then the following. First, we specify the initial potentials $\Phi = \Psi$.
Choosing the constant solution, we also have $\Phi' = 0$.
Next, we have to generate initial data for the particle list. Starting from a uniform particle distribution, the initial positions
of the particles can be assigned by an infinitesimal displacement, given by the gradient of a scalar function,
$\delta x^i = \delta^{ij} \zeta_{,j}$. The resulting ``bare'' mass density contrast will be $\delta = -\Delta \zeta$. The scalar
function $\zeta$ is then uniquely (up to an irrelevant homogeneous piece) determined by the linearized version of Eq.~(\ref{eq:phi}),
\be
\label{eq:displacement}
 \Delta \Phi - 3 \mathcal{H}^2 \Psi = 4 \pi G a^2 \bar{\rho} \left(-\Delta \zeta + 3 \Phi\right) \, .
\ee
This linear relation can be solved algebraically with the Fourier method. Equivalently, one could determine $\zeta$
using the relations (\ref{eq:Bardeen}) and the linear solutions (\ref{eq:linearsolutions}).

Assuming a negligible initial velocity dispersion, the initial peculiar velocities of the particles can be worked out from
Eqs.~(\ref{eq:Bardeen}) and (\ref{eq:linearsolutions}) as well. Ignoring again the decaying solution, one finds
\be
\label{eq:velocityic}
 \frac{\partial x^i}{\partial \tau} = - \frac{2}{3 \mathcal{H}} \delta^{ij} \Psi_{,j} \, .
\ee

In fact, Eqs.~(\ref{eq:displacement}) and (\ref{eq:velocityic}) can simply be regarded as the Zel'dovich approximation in
longitudinal gauge. It is worth noting the important difference to the corresponding approximation which is used to generate
initial data for Newtonian $N$-body simulations. In Newtonian gravity, the concept of a horizon is absent, and one wants
Eq.~(\ref{eq:poisson}) to hold on all scales. It is therefore reasonable to identify the initial density perturbation $\delta$
with another gauge-invariant measure thereof, namely the density perturbation in comoving gauge, which is denoted by $D$ in \cite{2008cmbg.book} and is related to
the gauge-invariant potential $\Phi$ precisely by Eq.~(\ref{eq:poisson}). The Newtonian potential $\psi_N$ is then again
identified with $\Phi = \Psi$. These identifications can be done consistently because there is an exact
correspondence between the Newtonian and relativistic solutions (taken in an appropriate gauge and assuming a pressureless equation
of state) at the level of scalar linear perturbations. This correspondence breaks down at the nonlinear level (or if the
pressureless assumption is not valid), and the Newtonian and relativistic solutions live in completely different worlds. Most
importantly, there is no particular gauge where Newtonian quantities can be identified with relativistic ones. All one can hope is
that there is an approximate correspondence which is reasonable for practical purposes, and this is the idea behind the
``dictionary'' which was proposed in \cite{Green:2011wc}.

Owing to the gauge-dependence of the density contrast $\delta$, the initial particle configuration for the same linear solution
(\ref{eq:linearsolutions}) is \textit{different} for a relativistic simulation compared to a Newtonian one. In particular, the
Newtonian Zel'dovich approximation simply has $\zeta = -2 \psi_N / 3 \mathcal{H}^2$, which corresponds to the relativistic
expression in comoving gauge. In the longitudinal gauge employed in our framework, $\zeta$ is given by Eq.~(\ref{eq:displacement}).
Evidently, on very long wavelengths where $k \ll \mathcal{H}$, we have that $\Delta\zeta \propto \Phi$ and thus
$\zeta$ can become large. This is, however, not a practical problem as the density contrast $\delta$ (and therefore the perturbation
of the particle number density) is of order $\Phi$ by construction and thus small. The long wavelength part of $\zeta_{,j}$ is just
a constant time-independent shift acting on a ``fictitious'' regular particle configuration that has no physical significance. In
particular, this configuration is never physically realized, even if one follows the particle trajectories backwards in time. Note
also that on super-horizon scales, where perturbations are always linear, one can always translate between different gauges by employing
a linear gauge transformation.

\bibliographystyle{utcaps}
\bibliography{julian}

\providecommand{\href}[2]{#2}\begingroup\raggedright\begin{thebibliography}{10}

\bibitem{Obradovic:2011mt}
M.~Obradovic, M.~Kunz, M.~Hindmarsh, and I.~T. Iliev, ``{Particle motion in
  weak relativistic gravitational fields},''
  \href{http://dx.doi.org/10.1103/PhysRevD.86.064018}{{\em Phys.Rev.}
  {\bfseries D86} (2012) 064018},
\href{http://arxiv.org/abs/1106.5866}{{\ttfamily arXiv:1106.5866
  [astro-ph.CO]}}.

\bibitem{Brax:2012ie}
P.~Brax, C.~Burrage, and A.-C. Davis, ``{Shining Light on Modifications of
  Gravity},'' \href{http://dx.doi.org/10.1088/1475-7516/2012/10/016}{{\em JCAP}
  {\bfseries 1210} (2012) 016},
\href{http://arxiv.org/abs/1206.1809}{{\ttfamily arXiv:1206.1809 [hep-th]}}.

\bibitem{Magueijo:1996px}
J.~Magueijo, A.~Albrecht, P.~Ferreira, and D.~Coulson, ``{The Structure of
  Doppler peaks induced by active perturbations},''
  \href{http://dx.doi.org/10.1103/PhysRevD.54.3727}{{\em Phys.Rev.} {\bfseries
  D54} (1996) 3727--3744},
\href{http://arxiv.org/abs/astro-ph/9605047}{{\ttfamily arXiv:astro-ph/9605047
  [astro-ph]}}.

\bibitem{Durrer:1997ep}
R.~Durrer and M.~Kunz, ``{Cosmic microwave background anisotropies from scaling
  seeds: Generic properties of the correlation functions},''
  \href{http://dx.doi.org/10.1103/PhysRevD.57.3199}{{\em Phys.Rev.} {\bfseries
  D57} (1998) R3199--R3203},
\href{http://arxiv.org/abs/astro-ph/9711133}{{\ttfamily arXiv:astro-ph/9711133
  [astro-ph]}}.

\bibitem{Saltas:2010tt}
I.~D. Saltas and M.~Kunz, ``{Anisotropic stress and stability in modified
  gravity models},'' \href{http://dx.doi.org/10.1103/PhysRevD.83.064042}{{\em
  Phys.Rev.} {\bfseries D83} (2011) 064042},
\href{http://arxiv.org/abs/1012.3171}{{\ttfamily arXiv:1012.3171 [gr-qc]}}.

\bibitem{Bonvin:2005ps}
C.~Bonvin, R.~Durrer, and M.~A. Gasparini, ``{Fluctuations of the luminosity
  distance},'' \href{http://dx.doi.org/10.1103/PhysRevD.85.029901,
  10.1103/PhysRevD.73.023523}{{\em Phys.Rev.} {\bfseries D73} (2006) 023523},
\href{http://arxiv.org/abs/astro-ph/0511183}{{\ttfamily arXiv:astro-ph/0511183
  [astro-ph]}}.

\bibitem{Bonvin:2011bg}
C.~Bonvin and R.~Durrer, ``{What galaxy surveys really measure},''
  \href{http://dx.doi.org/10.1103/PhysRevD.84.063505}{{\em Phys.Rev.}
  {\bfseries D84} (2011) 063505},
\href{http://arxiv.org/abs/1105.5280}{{\ttfamily arXiv:1105.5280
  [astro-ph.CO]}}.

\bibitem{Challinor:2011bk}
A.~Challinor and A.~Lewis, ``{The linear power spectrum of observed source
  number counts},'' \href{http://dx.doi.org/10.1103/PhysRevD.84.043516}{{\em
  Phys.Rev.} {\bfseries D84} (2011) 043516},
\href{http://arxiv.org/abs/1105.5292}{{\ttfamily arXiv:1105.5292
  [astro-ph.CO]}}.

\bibitem{Bertacca:2012tp}
D.~Bertacca, R.~Maartens, A.~Raccanelli, and C.~Clarkson, ``{Beyond the
  plane-parallel and Newtonian approach: Wide-angle redshift distortions and
  convergence in general relativity},''
  \href{http://dx.doi.org/10.1088/1475-7516/2012/10/025}{{\em JCAP} {\bfseries
  1210} (2012) 025},
\href{http://arxiv.org/abs/1205.5221}{{\ttfamily arXiv:1205.5221
  [astro-ph.CO]}}.

\bibitem{Umeh:2012pn}
O.~Umeh, C.~Clarkson, and R.~Maartens, ``{Nonlinear general relativistic
  corrections to redshift space distortions, gravitational lensing
  magnification and cosmological distances},''
\href{http://arxiv.org/abs/1207.2109}{{\ttfamily arXiv:1207.2109
  [astro-ph.CO]}}.

\bibitem{BenDayan:2012wi}
I.~Ben-Dayan, G.~Marozzi, F.~Nugier, and G.~Veneziano, ``{The second-order
  luminosity-redshift relation in a generic inhomogeneous cosmology},''
  \href{http://dx.doi.org/10.1088/1475-7516/2012/11/045}{{\em JCAP} {\bfseries
  1211} (2012) 045},
\href{http://arxiv.org/abs/1209.4326}{{\ttfamily arXiv:1209.4326
  [astro-ph.CO]}}.

\bibitem{Bonvin:2006en}
C.~Bonvin, R.~Durrer, and M.~Kunz, ``{The dipole of the luminosity distance: a
  direct measure of h(z)},''
  \href{http://dx.doi.org/10.1103/PhysRevLett.96.191302}{{\em Phys.Rev.Lett.}
  {\bfseries 96} (2006) 191302},
\href{http://arxiv.org/abs/astro-ph/0603240}{{\ttfamily arXiv:astro-ph/0603240
  [astro-ph]}}.

\bibitem{Abate:2008au}
A.~Abate and O.~Lahav, ``{The Three Faces of Omega\_m: Testing Gravity with Low
  and High Redshift SN Ia Surveys},''
  \href{http://dx.doi.org/10.1111/j.1745-3933.2008.00519.x}{{\em
  Mon.Not.Roy.Astron.Soc.} {\bfseries 389} (2008) L47--L51},
\href{http://arxiv.org/abs/0805.3160}{{\ttfamily arXiv:0805.3160 [astro-ph]}}.

\bibitem{Valkenburg:2013qwa}
W.~Valkenburg, M.~Kunz, and V.~Marra, ``{Intrinsic uncertainty on the nature of
  dark energy},''
\href{http://arxiv.org/abs/1302.6588}{{\ttfamily arXiv:1302.6588
  [astro-ph.CO]}}.

\bibitem{BenDayan:2013gc}
I.~Ben-Dayan, M.~Gasperini, G.~Marozzi, F.~Nugier, and G.~Veneziano, ``{Average
  and dispersion of the luminosity-redshift relation in the concordance
  model},''
\href{http://arxiv.org/abs/1302.0740}{{\ttfamily arXiv:1302.0740
  [astro-ph.CO]}}.

\bibitem{Marra:2013rba}
V.~Marra, L.~Amendola, I.~Sawicki, and W.~Valkenburg, ``{Cosmic variance and
  the measurement of the local Hubble parameter},''
\href{http://arxiv.org/abs/1303.3121}{{\ttfamily arXiv:1303.3121
  [astro-ph.CO]}}.

\bibitem{Fleury:2013uqa}
P.~Fleury, H.~Dupuy, and J.-P. Uzan, ``{Can all cosmological observations be
  accurately interpreted with a unique geometry?},''
\href{http://arxiv.org/abs/1304.7791}{{\ttfamily arXiv:1304.7791
  [astro-ph.CO]}}.

\bibitem{Rasanen:2006kp}
S.~R{\"{a}}s{\"{a}}nen, ``{Accelerated expansion from structure formation},''
  \href{http://dx.doi.org/10.1088/1475-7516/2006/11/003}{{\em JCAP} {\bfseries
  0611} (2006) 003},
\href{http://arxiv.org/abs/astro-ph/0607626}{{\ttfamily arXiv:astro-ph/0607626
  [astro-ph]}}.

\bibitem{Buchert:2007ik}
T.~Buchert, ``{Dark Energy from Structure: A Status Report},''
  \href{http://dx.doi.org/10.1007/s10714-007-0554-8}{{\em Gen.Rel.Grav.}
  {\bfseries 40} (2008) 467--527},
\href{http://arxiv.org/abs/0707.2153}{{\ttfamily arXiv:0707.2153 [gr-qc]}}.

\bibitem{Rasanen:2011ki}
S.~R{\"{a}}sa{\"{a}}en, ``{Backreaction: directions of progress},''
  \href{http://dx.doi.org/10.1088/0264-9381/28/16/164008}{{\em
  Class.Quant.Grav.} {\bfseries 28} (2011) 164008},
\href{http://arxiv.org/abs/1102.0408}{{\ttfamily arXiv:1102.0408
  [astro-ph.CO]}}.

\bibitem{Clarkson:2011zq}
C.~Clarkson, G.~Ellis, J.~Larena, and O.~Umeh, ``{Does the growth of structure
  affect our dynamical models of the universe? The averaging, backreaction and
  fitting problems in cosmology},''
  \href{http://dx.doi.org/10.1088/0034-4885/74/11/112901}{{\em Rept.Prog.Phys.}
  {\bfseries 74} (2011) 112901},
\href{http://arxiv.org/abs/1109.2314}{{\ttfamily arXiv:1109.2314
  [astro-ph.CO]}}.

\bibitem{Buchert:2011sx}
T.~Buchert and S.~R{\"{a}}s{\"{a}}nen, ``{Backreaction in late-time
  cosmology},''
  \href{http://dx.doi.org/10.1146/annurev.nucl.012809.104435}{{\em
  Ann.Rev.Nucl.Part.Sci.} {\bfseries 62} (2012) 57--79},
\href{http://arxiv.org/abs/1112.5335}{{\ttfamily arXiv:1112.5335
  [astro-ph.CO]}}.

\bibitem{Rasanen:2010wz}
S.~R{\"{a}}s{\"{a}}nen, ``{Applicability of the linearly perturbed FRW metric
  and Newtonian cosmology},''
  \href{http://dx.doi.org/10.1103/PhysRevD.81.103512}{{\em Phys.Rev.}
  {\bfseries D81} (2010) 103512},
\href{http://arxiv.org/abs/1002.4779}{{\ttfamily arXiv:1002.4779
  [astro-ph.CO]}}.

\bibitem{lrr-2011-6}
M.~Shibata and K.~Taniguchi, ``Coalescence of Black Hole-Neutron Star
  Binaries,'' \href{http://dx.doi.org/10.12942/lrr-2011-6}{{\em Living Reviews
  in Relativity} {\bfseries 14} no.~6, (2011) }.

\bibitem{Green:2010qy}
S.~R. Green and R.~M. Wald, ``{A new framework for analyzing the effects of
  small scale inhomogeneities in cosmology},''
  \href{http://dx.doi.org/10.1103/PhysRevD.83.084020}{{\em Phys.Rev.}
  {\bfseries D83} (2011) 084020},
\href{http://arxiv.org/abs/1011.4920}{{\ttfamily arXiv:1011.4920 [gr-qc]}}.

\bibitem{Green:2011wc}
S.~R. Green and R.~M. Wald, ``{Newtonian and Relativistic Cosmologies},''
  \href{http://dx.doi.org/10.1103/PhysRevD.85.063512}{{\em Phys.Rev.}
  {\bfseries D85} (2012) 063512},
\href{http://arxiv.org/abs/1111.2997}{{\ttfamily arXiv:1111.2997 [gr-qc]}}.

\bibitem{Rampf:2013ewa}
C.~Rampf and G.~Rigopoulos, ``{Initial conditions for cold dark matter
  particles and General Relativity},''
  \href{http://dx.doi.org/10.1103/PhysRevD.87.123525}{{\em Phys.Rev.}
  {\bfseries D87} (2013) 123525},
\href{http://arxiv.org/abs/1305.0010}{{\ttfamily arXiv:1305.0010
  [astro-ph.CO]}}.

\bibitem{Rigopoulos:2013nda}
G.~Rigopoulos and W.~Valkenburg, ``{On the accuracy of N-body simulations at
  very large scales},''
\href{http://arxiv.org/abs/1308.0057}{{\ttfamily arXiv:1308.0057
  [astro-ph.CO]}}.

\bibitem{Adamek:2014qja}
J.~Adamek, E.~Di~Dio, R.~Durrer, and M.~Kunz, ``{The distance-redshift relation
  in plane symmetric universes},''
\href{http://arxiv.org/abs/1401.3634}{{\ttfamily arXiv:1401.3634
  [astro-ph.CO]}}.

\bibitem{Brustein:1994kn}
R.~Brustein, M.~Gasperini, M.~Giovannini, V.~F. Mukhanov, and G.~Veneziano,
  ``{Metric perturbations in dilaton driven inflation},''
  \href{http://dx.doi.org/10.1103/PhysRevD.51.6744}{{\em Phys.Rev.} {\bfseries
  D51} (1995) 6744--6756},
\href{http://arxiv.org/abs/hep-th/9501066}{{\ttfamily arXiv:hep-th/9501066
  [hep-th]}}.

\bibitem{Cartier:2003jz}
C.~Cartier, R.~Durrer, and E.~J. Copeland, ``{Cosmological perturbations and
  the transition from contraction to expansion},''
  \href{http://dx.doi.org/10.1103/PhysRevD.67.103517}{{\em Phys.Rev.}
  {\bfseries D67} (2003) 103517},
\href{http://arxiv.org/abs/hep-th/0301198}{{\ttfamily arXiv:hep-th/0301198
  [hep-th]}}.

\bibitem{Turnbull:2011ty}
S.~J. Turnbull, M.~J. Hudson, H.~A. Feldman, M.~Hicken, R.~P. Kirshner, {\em
  et~al.}, ``{Cosmic flows in the nearby universe from Type Ia Supernovae},''
  \href{http://dx.doi.org/10.1111/j.1365-2966.2011.20050.x}{{\em
  Mon.Not.Roy.Astron.Soc.} {\bfseries 420} (2012) 447--454},
\href{http://arxiv.org/abs/1111.0631}{{\ttfamily arXiv:1111.0631
  [astro-ph.CO]}}.

\bibitem{Efstathiou:1985re}
G.~Efstathiou, M.~Davis, C.~Frenk, and S.~D. White, ``{Numerical Techniques for
  Large Cosmological N-Body Simulations},''
\href{http://dx.doi.org/10.1086/191003}{{\em Astrophys.J.Suppl.} {\bfseries 57}
  (1985) 241--260}.

\bibitem{Nbody}
S.~J. Aarseth, {\em Gravitational N-body Simulations: Tools and Algorithms}.
\newblock Cambridge University Press, 2003.

\bibitem{Springel:2005nw}
V.~Springel, S.~D. White, A.~Jenkins, C.~S. Frenk, N.~Yoshida, {\em et~al.},
  ``{Simulating the joint evolution of quasars, galaxies and their large-scale
  distribution},'' \href{http://dx.doi.org/10.1038/nature03597}{{\em Nature}
  {\bfseries 435} (2005) 629--636},
\href{http://arxiv.org/abs/astro-ph/0504097}{{\ttfamily arXiv:astro-ph/0504097
  [astro-ph]}}.

\bibitem{Springel:2008cc}
V.~Springel, J.~Wang, M.~Vogelsberger, A.~Ludlow, A.~Jenkins, {\em et~al.},
  ``{The Aquarius Project: the subhalos of galactic halos},''
  \href{http://dx.doi.org/10.1111/j.1365-2966.2008.14066.x}{{\em
  Mon.Not.Roy.Astron.Soc.} {\bfseries 391} (2008) 1685--1711},
\href{http://arxiv.org/abs/0809.0898}{{\ttfamily arXiv:0809.0898 [astro-ph]}}.

\bibitem{Matsubara:2000pr}
T.~Matsubara, ``{The gravitational lensing in redshift-space correlation
  functions of galaxies and quasars},''
  \href{http://dx.doi.org/10.1086/312762}{{\em Astrophys.J.Lett.} {\bfseries
  537} (2000) L77},
\href{http://arxiv.org/abs/astro-ph/0004392}{{\ttfamily arXiv:astro-ph/0004392
  [astro-ph]}}.

\bibitem{Chisari:2011iq}
N.~E. Chisari and M.~Zaldarriaga, ``{Connection between Newtonian simulations
  and general relativity},''
  \href{http://dx.doi.org/10.1103/PhysRevD.84.089901,
  10.1103/PhysRevD.83.123505}{{\em Phys.Rev.} {\bfseries D83} (2011) 123505},
\href{http://arxiv.org/abs/1101.3555}{{\ttfamily arXiv:1101.3555
  [astro-ph.CO]}}.

\bibitem{Colberg:2000zv}
{\bfseries VIRGO} Collaboration, J.~Colberg {\em et~al.}, ``{Clustering of
  galaxy clusters in CDM universes},''
  \href{http://dx.doi.org/10.1046/j.1365-8711.2000.03832.x}{{\em
  Mon.Not.Roy.Astron.Soc.} {\bfseries 319} (2000) 209},
\href{http://arxiv.org/abs/astro-ph/0005259}{{\ttfamily arXiv:astro-ph/0005259
  [astro-ph]}}.

\bibitem{Evrard:2001hu}
{\bfseries VIRGO} Collaboration, A.~Evrard {\em et~al.}, ``{Galaxy clusters in
  Hubble volume simulations: Cosmological constraints from sky survey
  populations},'' \href{http://dx.doi.org/10.1086/340551}{{\em Astrophys.J.}
  {\bfseries 573} (2002) 7--36},
\href{http://arxiv.org/abs/astro-ph/0110246}{{\ttfamily arXiv:astro-ph/0110246
  [astro-ph]}}.

\bibitem{Gottlober:2006sx}
S.~Gottlober, G.~Yepes, A.~Khalatyan, R.~Sevilla, and V.~Turchaninov, ``{Dark
  and baryonic matter in the MareNostrum Universe},''
  \href{http://dx.doi.org/10.1063/1.2409061}{{\em AIP Conf.Proc.} {\bfseries
  878} (2006) 3--9},
\href{http://arxiv.org/abs/astro-ph/0610622}{{\ttfamily arXiv:astro-ph/0610622
  [astro-ph]}}.

\bibitem{Park:2005ek}
C.~Park, J.~Kim, and J.~R. Gott~III, ``{Effects of gravitational evolution,
  biasing, and redshift space distortion on topology},''
  \href{http://dx.doi.org/10.1086/452621}{{\em Astrophys.J.} {\bfseries 633}
  (2005) 1--10},
\href{http://arxiv.org/abs/astro-ph/0503584}{{\ttfamily arXiv:astro-ph/0503584
  [astro-ph]}}.

\bibitem{NumericalRecipes}
W.~H. Press, S.~A. Teukolsky, W.~T. Vetterling, and B.~P. Flannery, {\em
  Numerical Recipes}, ch.~20.
\newblock Cambridge University Press, 3rd~ed., 2007.

\bibitem{Douglas1956}
J.~Douglas and H.~H. Rachford, ``On the Numerical Solution of Heat Conduction
  Problems in Two and Three Space Variables,''
  \href{http://dx.doi.org/10.1090/S0002-9947-1956-0084194-4}{{\em
  Trans.Amer.Math.Soc.} {\bfseries 82} (1956) 421--439}.

\bibitem{Douglas1962a}
J.~Douglas, ``Alternating direction methods for three space variables,''
  \href{http://dx.doi.org/10.1007/BF01386295}{{\em Num.Math.} {\bfseries 4}
  (1962) 41--63}.

\bibitem{MattorWH95}
N.~Mattor, T.~J. Williams, and D.~W. Hewett, ``{Algorithm for Solving
  Tridiagonal Matrix Problems in Parallel},''
  \href{http://dx.doi.org/10.1016/0167-8191%2895%2900033-0}{{\em Parallel
  Computing} {\bfseries 21} (1995) 1769--1782}.

\bibitem{Chan:2009ew}
K.~Chan and R.~Scoccimarro, ``{Large-Scale Structure in Brane-Induced Gravity
  II. Numerical Simulations},''
  \href{http://dx.doi.org/10.1103/PhysRevD.80.104005}{{\em Phys.Rev.}
  {\bfseries D80} (2009) 104005},
\href{http://arxiv.org/abs/0906.4548}{{\ttfamily arXiv:0906.4548
  [astro-ph.CO]}}.

\bibitem{Li:2011vk}
B.~Li, G.-B. Zhao, R.~Teyssier, and K.~Koyama, ``{ECOSMOG: An Efficient Code
  for Simulating Modified Gravity},''
  \href{http://dx.doi.org/10.1088/1475-7516/2012/01/051}{{\em JCAP} {\bfseries
  1201} (2012) 051},
\href{http://arxiv.org/abs/1110.1379}{{\ttfamily arXiv:1110.1379
  [astro-ph.CO]}}.

\bibitem{Lee:2012bm}
J.~Lee, G.-B. Zhao, B.~Li, and K.~Koyama, ``{Modified Gravity Spins Up Galactic
  Halos},'' \href{http://dx.doi.org/10.1088/0004-637X/763/1/28}{{\em
  Astrophys.J.} {\bfseries 763} (2013) 28},
\href{http://arxiv.org/abs/1204.6608}{{\ttfamily arXiv:1204.6608
  [astro-ph.CO]}}.

\bibitem{Brax:2013mua}
P.~Brax, A.-C. Davis, B.~Li, H.~A. Winther, and G.-B. Zhao, ``{Systematic
  simulations of modified gravity: chameleon models},''
  \href{http://dx.doi.org/10.1088/1475-7516/2013/04/029}{{\em JCAP} {\bfseries
  1304} (2013) 029},
\href{http://arxiv.org/abs/1303.0007}{{\ttfamily arXiv:1303.0007
  [astro-ph.CO]}}.

\bibitem{Puchwein:2013lza}
E.~Puchwein, M.~Baldi, and V.~Springel, ``{Modified Gravity-GADGET: A new code
  for cosmological hydrodynamical simulations of modified gravity models},''
\href{http://arxiv.org/abs/1305.2418}{{\ttfamily arXiv:1305.2418
  [astro-ph.CO]}}.

\bibitem{Ballesteros:2011cm}
G.~Ballesteros, L.~Hollenstein, R.~K. Jain, and M.~Kunz, ``{Nonlinear
  cosmological consistency relations and effective matter stresses},''
  \href{http://dx.doi.org/10.1088/1475-7516/2012/05/038}{{\em JCAP} {\bfseries
  1205} (2012) 038},
\href{http://arxiv.org/abs/1112.4837}{{\ttfamily arXiv:1112.4837
  [astro-ph.CO]}}.

\bibitem{Bruni:2013mua}
M.~Bruni, D.~B. Thomas, and D.~Wands, ``{Computing General Relativistic effects
  from Newtonian N-body simulations: Frame dragging in the post-Friedmann
  approach},''
\href{http://arxiv.org/abs/1306.1562}{{\ttfamily arXiv:1306.1562
  [astro-ph.CO]}}.

\bibitem{HockneyEastwood}
R.~W. Hockney and J.~W. Eastwood, {\em Computer Simulation Using Particles}.
\newblock Institute of Physics Publ., 1999.

\bibitem{Crocce:2006ve}
M.~Crocce, S.~Pueblas, and R.~Scoccimarro, ``{Transients from Initial
  Conditions in Cosmological Simulations},''
  \href{http://dx.doi.org/10.1111/j.1365-2966.2006.11040.x}{{\em
  Mon.Not.Roy.Astron.Soc.} {\bfseries 373} (2006) 369--381},
\href{http://arxiv.org/abs/astro-ph/0606505}{{\ttfamily arXiv:astro-ph/0606505
  [astro-ph]}}.

\bibitem{Bardeen:1980kt}
J.~M. Bardeen, ``{Gauge Invariant Cosmological Perturbations},''
\href{http://dx.doi.org/10.1103/PhysRevD.22.1882}{{\em Phys.Rev.} {\bfseries
  D22} (1980) 1882--1905}.

\bibitem{2008cmbg.book}
R.~{Durrer}, {\em {The Cosmic Microwave Background}}.
\newblock Cambridge University Press, 2008.

\end{thebibliography}\endgroup

\end{document}